\pdfoutput=1

\documentclass[11pt,twoside,a4paper,cmspaper,final,collab]{cms-tdr}
\def\svnVersion{cb47fa7-D}\def\svnDate{2020/03/13}\def\cmsCernNoTag{CERN-EP-2019-107}\def\cmsCernDate{\today}\def\cmsMessage{"Published in the European Physical Journal C as \href{http://dx.doi.org/10.1140/epjc/s10052-020-7773-5}{\doi{10.1140/epjc/s10052-020-7773-5}.}"}

\begin{document}\cmsNoteHeader{B2G-18-002}

\hyphenation{had-ron-i-za-tion}
\hyphenation{cal-or-i-me-ter}
\hyphenation{de-vices}
\RCS$HeadURL$
\RCS$Id$

\newlength\cmsFigWidth
\newlength\cmsTabSkip\setlength{\cmsTabSkip}{1ex}
\providecommand{\cmsTable}[1]{\resizebox{\textwidth}{!}{#1}}
\providecommand{\NA}{\ensuremath{\text{---}}}

\ifthenelse{\boolean{cms@external}}{\setlength\cmsFigWidth{0.49\textwidth}}{\setlength\cmsFigWidth{0.65\textwidth}}
\ifthenelse{\boolean{cms@external}}{\providecommand{\cmsLeft}{upper\xspace}}{\providecommand{\cmsLeft}{left\xspace}}
\ifthenelse{\boolean{cms@external}}{\providecommand{\cmsRight}{lower\xspace}}{\providecommand{\cmsRight}{right\xspace}}

\newcommand{\INTLUMI}     {77.3\xspace}
\newcommand{\MassExclWPr}{3.8\xspace}
\newcommand{\MassExclZPr}{3.5\xspace}
\newcommand{\BulkGMassMin}{1.2\xspace}
\newcommand{\BulkGMassMinXsec}{27\xspace}
\newcommand{\BulkGMassMax}{5.2\xspace}
\newcommand{\BulkGWWMassMinXsec}{20\xspace}
\newcommand{\BulkGWWMassMaxXsec}{0.2\xspace}
\newcommand{\BulkGZZMassMinXsec}{27\xspace}
\newcommand{\Zo}{\cPZ \xspace}
\newcommand{\mjj}{\ensuremath{m_\mathrm{jj}}}
\newcommand{\mJ}{\ensuremath{m_{\text{jet}}}}
\newcommand{\nsubj}{\ensuremath{\tau_{21}}\xspace}
\newcommand{\nsubjDDT}{\ensuremath{\tau_{21}^\text{DDT}}\xspace}
\newcommand{\BulkG}{\ensuremath{\PXXG_{\text{bulk}}}\xspace}
\newcommand\T{\rule[-.50em]{0pt}{1.50em}}
\newcommand\Y{\rule[-.0em]{0pt}{1.00em}}
\newcommand{\MVV}{\ensuremath{m_\mathrm{jj}}\xspace}
\newcommand{\MJ}{\ensuremath{m_\text{jet}}\xspace}
\newcommand{\MJO}{\ensuremath{m_\text{jet1}}\xspace}
\newcommand{\MJT}{\ensuremath{m_\text{jet2}}\xspace}
\newcommand{\PTk}{\ensuremath{p_{\mathrm{T},k}}\xspace}
\newcommand{\MX}{\ensuremath{m_\mathrm{X}}\xspace}
\newcolumntype{X}{D{+}{\,\pm\,}{5,3}}

\cmsNoteHeader{B2G-18-002}
\title{A multi-dimensional search for new heavy resonances decaying to boosted \PW{}\PW{}, \PW{}\PZ{}, or \PZ{}\PZ boson pairs in the dijet final state at 13\TeV}
\titlerunning{A multi-dimensional search for heavy resonances decaying to diboson pairs}

\date{\today}

\abstract{
A search in an all-jet final state for new massive resonances decaying to \PW{}\PW, \PW{}\PZ, or \PZ{}\PZ boson pairs using a novel analysis method is presented. The analysis is performed on data corresponding to an integrated luminosity of \INTLUMI{}\fbinv recorded with the CMS experiment at the LHC at a centre-of-mass energy of 13\TeV.
The search is focussed on potential narrow-width resonances with masses above 1.2\TeV, where the decay products of each \PW\ or \PZ\ boson are expected to be collimated into a single, large-radius jet. The signal is extracted using a three-dimensional maximum likelihood fit of the two jet masses and the dijet invariant mass, yielding an improvement in sensitivity of up to 30\% relative to previous search methods. No excess is observed above the estimated standard model background. In a heavy vector triplet model, spin-1 \PZpr{} and \PWpr{} resonances with masses below \MassExclZPr{} and \MassExclWPr{}\TeV, respectively, are excluded at 95\% confidence level.
In a bulk graviton model, upper limits on cross sections are set between \BulkGMassMinXsec and \BulkGWWMassMaxXsec{}\unit{fb} for resonance masses between \BulkGMassMin and \BulkGMassMax{}\TeV, respectively. The limits presented in this paper are the best to date in the dijet final state. }

\hypersetup{%
pdfauthor={CMS Collaboration},%
pdftitle={A multidimensional search for new heavy resonances decaying to boosted WW, WZ or ZZ boson pairs in the dijet final state},%
pdfsubject={CMS},%
pdfkeywords={CMS, physics, diboson resonances, substructure}}

\maketitle

\section{Introduction}
\label{sec:Introduction}

The standard model (SM) of particle physics has been exceptionally successful in accommodating a multitude of experimental measurements and observations, yet it falls short in a variety of aspects.
These shortcomings motivate theoretical extensions of the SM that typically introduce new particles, which could be created in proton-proton (\Pp{}\Pp) collisions at the CERN LHC. In this analysis, we test theoretical models that predict new heavy resonances that decay to pairs of vector bosons (\PW{} and \PZ{} bosons, collectively referred to as \PV bosons).
These models usually aim to clarify open questions in the SM such as the large difference between the electroweak and the Planck scales.
We consider the bulk scenario of the Randall--Sundrum (RS) model with warped extra dimensions~\cite{Randall:1999ee,Randall:1999vf,Agashe:2007zd, Fitzpatrick:2007qr, Antipin:2007pi}, where the spin-2 bulk graviton has an enhanced branching fraction to massive particles, and the heavy vector triplet (HVT) framework~\cite{Pappadopulo:2014qza}, which serves as a template that reproduces a large class of explicit models predicting \mbox{spin-1} resonances.

No significant deviations from the SM background expectation have been observed in previous searches by the CMS Collaboration for such particles in the \PV{}\PV{}~\cite{Khachatryan:2014hpa,Khachatryan:2014gha,Khachatryan:2014xja,Sirunyan:2016cao,Sirunyan:2018iff} and \PV{}\PH{}~\cite{Khachatryan:2015ywa,Sirunyan:2017wto,Khachatryan:2016cfx,Khachatryan:2015bma,Khachatryan:2016yji,Khachatryan:2016cfa} channels, where \PH{} denotes the Higgs boson. Similar results were obtained independently by the ATLAS Collaboration in \PV{}\PV{}~\cite{ATLASwprimeWZPAS,Aad:2014xka,Aad:2015ufa,Aaboud:2017ahz,Aaboud:2017eta,Aaboud:2016okv} and \PV{}\PH{}~\cite{Aad:2015yza,ATLASVH13,Aad:2015owa} resonance searches. In addition, statistical combinations of diboson and leptonic decay channels of the 2016 data set~\cite{Sirunyan:2019vgt,Aaboud:2018bun} have been performed, which extend the exclusion regions of the individual analyses.
Lower limits on the masses of these resonances have been set at the \TeVns{} scale.
The search presented here focusses on resonances with masses above 1.2\TeV, in the decays of which the vector bosons are produced at high Lorentz boost. Because of the large boost of the vector bosons, their decay products are merged into single, large-radius jets, leading to dijet final states. These jets are identified through dedicated jet substructure algorithms.
Compared to previous analyses in this final state~\cite{Khachatryan:2014hpa,Sirunyan:2016cao,Sirunyan:2017acf,Aad:2015owa,Aaboud:2017eta,Aaboud:2016okv}, an improved background estimation and signal extraction procedure based on a three-dimensional~(3D) maximum likelihood fit is employed, increasing the sensitivity of the analysis.
The method can be applied to any search with final states expected to cause resonant behaviour in three observables, whereas previous methods used solely the invariant mass of the final decay products as the search variable.
The improved sensitivity and scope has motivated a reanalysis of the \Pp{}\Pp{} collision data collected by the CMS experiment during the 2016 data taking period, as well as a first analysis of the 2017 data set, corresponding to a total integrated luminosity of  \INTLUMI{}\fbinv at a centre-of-mass energy of 13\TeV.

\section{The CMS detector}
\label{sec:cmsdetector}

The central feature of the CMS apparatus is a superconducting solenoid of 6\unit{m} internal diameter, providing a magnetic field of 3.8\unit{T}. Within the solenoid volume are a silicon pixel and strip tracker, a lead tungstate crystal electromagnetic calorimeter, and a brass and scintillator hadron calorimeter, each composed of a barrel and two endcap sections. Forward calorimeters extend the pseudorapidity ($\eta$) coverage provided by the barrel and endcap detectors. Muons are detected in gas-ionization chambers embedded in the steel flux-return yoke outside the solenoid. A more detailed description of the CMS detector, together with a definition of the coordinate system used and the relevant kinematic variables, can be found in Ref.~\cite{Chatrchyan:2008zzk}.

Events of interest are selected using a two-tiered trigger system~\cite{Khachatryan:2016bia}. The first level, composed of custom hardware processors, uses information from the calorimeters and muon detectors to select events at a rate of around 100\unit{kHz} within a time interval of less than 4\mus. The second level, known as the high-level trigger, consists of a farm of processors running a version of the full event reconstruction software optimized for fast processing, and reduces the event rate to around 1\unit{kHz} before data storage.

\section{Simulated events}
\label{sec:samples}

The resonances associated with the considered phenomenologies are the bulk gravitons (\BulkG) generated for the bulk scenario~\cite{Agashe:2007zd, Fitzpatrick:2007qr, Antipin:2007pi} of the RS model of warped extra dimensions~\cite{Randall:1999ee,Randall:1999vf}, and the heavy new bosons (\PWpr and \PZpr) that can be part of an heavy vector triplet~\cite{Pappadopulo:2014qza} or can be mass degenerate as a vector singlet~\cite{Grojean:2011vu,Salvioni:2009mt}.

The bulk graviton model is characterized by two free parameters: the mass of the first Kaluza--Klein (KK) excitation of a spin-2 boson (the KK bulk graviton), and the ratio $\tilde{\kappa}=\kappa\sqrt{8\pi}/\Mpl$, with $\kappa$ being the unknown curvature scale of the extra dimension and \Mpl the Planck mass. A scenario with $\tilde{\kappa}=0.5$ is considered in this analysis, as motivated in Ref.~\cite{Oliveira:2014kla}.

The HVT framework generically represents a large number of models predicting additional gauge bosons, such as the composite Higgs~\cite{Bellazzini:2014yua,CHM2,Composite2,Greco:2014aza,Lane:2016kvg} and little Higgs~\cite{Schmaltz:2005ky,ArkaniHamed:2002qy} models. The benchmark points are formulated in terms of a few parameters: two coefficients $c_\mathrm{F}$ and $c_{\PH{}}$, that scale the couplings of the additional gauge bosons to fermions; to the Higgs boson and longitudinally polarized SM vector bosons, respectively, and $g_\mathrm{V}$, representing the typical strength of the new vector boson interaction.
For the analysis presented here, samples were simulated in the HVT model B, corresponding to $g_\mathrm{V}=3$, $c_{\PH{}}=-0.98$, and $c_\mathrm{F}=1.02$~\cite{Pappadopulo:2014qza}. For these parameters, the new resonances are narrow and have large branching fractions to vector boson pairs, while the fermionic couplings are suppressed.

All signals considered in the analysis satisfy the narrow-width approximation. The quoted results are therefore valid independent of the exact theoretical signal widths as long as the resonance widths remain smaller than the detector resolution.
This makes our modelling of the detector effects on the signal shape independent of the actual model used for generating the events. All simulated samples are produced with a relative resonance width of 0.1\%, in order to be within the validity of the narrow-width approximation. Monte Carlo (MC) simulated events of the bulk graviton and HVT signals are generated at leading-order (LO) in quantum chromodynamics (QCD) with \MGvATNLO versions 2.2.2 and 2.4.3~\cite{Alwall:2014hca} and hadronization showering is simulated with \PYTHIA{} versions 8.205 and 8.230~\cite{Sjostrand:2014zea}, for 2016 and 2017 detector conditions, respectively.
The NNPDF 3.0~\cite{Ball:2014uwa} LO parton distribution functions (PDFs) are used together with the CUETP8M1~\cite{Khachatryan:2015pea} and CP5~\cite{CMS-GEN-17-001} underlying event tunes in \PYTHIA{} for 2016 and 2017 conditions, respectively.

{\tolerance=800 Simulated samples of the SM background processes are used to optimize the analysis and create background templates, as described in Section~\ref{sec:multidimfit}.
The QCD multijet production is simulated with four generator configurations: \PYTHIA{} only, the LO mode of \MGvATNLO{}~\cite{Alwall:2007fs} matched and showered with \PYTHIA{}, \POWHEG{}~\cite{Alioli:2010xa,Nason:2004rx,Frixione:2007vw,Alioli:2010xd} matched and showered with \PYTHIA{}, and \HERWIG{++}~2.7.1 \cite{Bahr:2008pv} with the CUETHS1 tune~\cite{Khachatryan:2015pea}.
Top quark pair production is modelled at next-to-LO (NLO) with \POWHEG{}~\cite{Alioli:2011as}, showered with \PYTHIA{}. To calculate systematic uncertainties related to the vector boson tagging efficiency, two additional simulated samples of top quark production at LO are used: one generated with \MGvATNLO{} and interfaced with \PYTHIA{}, and the second one generated and showered with \PYTHIA{}.
The production of \PW{}+jets and \PZ{}+jets (\PV{}+jets) is simulated at LO with \MGvATNLO{} matched and showered with \PYTHIA{}. The same underlying event tune as for the signal samples is used for those of the background. Two corrections dependent on the transverse momentum (\PT{})~\cite{Kallweit:2014xda,Kallweit:2015dum} are applied to the \PV{}+jets backgrounds to correct the \PT-distribution of the vector bosons computed at LO in QCD to the one predicted at NLO in QCD, and to account for electroweak effects at high \PT. The NNPDF 3.1~\cite{Ball:2017nwa} next-to-NLO (NNLO) PDFs are employed for simulated \PV{}+jets events with the 2017 data taking conditions for both the 2016 and 2017 data analyses.\par}

All samples are processed through a \GEANTfour-based~\cite{Agostinelli:2002hh} simulation of the CMS detector. To simulate the effect of additional pp collisions within the same or adjacent bunch crossings (pileup), additional inelastic events are generated using \PYTHIA{} and superimposed on the hard-scattering events. The MC simulated events are weighted to reproduce the distribution of the number of reconstructed pileup vertices observed in the 2016 and 2017 data separately.

\section{Reconstruction and selection of events}
\label{sec:eventreconstruction}

\subsection{Jet reconstruction}
\label{sec:jetreconstruction}
Event reconstruction is based on the particle flow (PF) algorithm~\cite{Sirunyan:2017ulk}, which reconstructs and identifies individual particles with information from the various elements of the CMS detector. Jets are reconstructed from these particles, using the anti-\kt{} jet clustering algorithm~\cite{Cacciari:2008gp} with a distance parameter of $R=0.8$ (AK8 jets) as implemented in the \FASTJET{} package~\cite{Cacciari:2011ma}.
In order to mitigate the effect of pileup, two different algorithms are used: for 2016 data and simulation, charged particles identified as originating from pileup vertices are excluded before jet clustering begins. For 2017, we take advantage of the pileup per particle identification (PUPPI)~\cite{Bertolini:2014bba} algorithm. This method uses local shape information of charged pileup, event pileup properties, and tracking information in order to rescale the four-momentum of each neutral and charged PF candidate with a weight that describes the likelihood that each particle originates from a pileup interaction. All jets are further required to pass tight jet identification requirements~\cite{CMS-PAS-JME-16-003}. Jets are corrected for nonlinearities in \PT{} and $\eta$ using jet energy corrections as described in Ref.~\cite{Khachatryan:2016kdb}. Additionally, residual contributions from pileup are corrected using the approach outlined in Ref.~\cite{jetarea_fastjet}.

Two variables are used to tag jets as potentially originating from vector boson decays to quarks for further event selection: the ``groomed" mass of the jet obtained using a modified mass drop algorithm~\cite{Dasgupta:2013ihk,Butterworth:2008iy} known as soft drop~\cite{Larkoski:2014wba}, and the $N$-subjettiness ratio $\nsubj=\tau_{2}/\tau_{1}$ obtained with the $N$-subjettiness algorithm~\cite{nsubjettiness}. For both 2016 and 2017 data these observables are reconstructed from AK8 jets with PUPPI pileup mitigation applied, decreasing their dependence on pileup as shown in Ref.~\cite{CMS-PAS-JME-16-003}, while the overall jet four-momenta are calculated using the pileup mitigation algorithms as described above.

The groomed jet mass is calculated using the soft drop algorithm, with angular exponent $\beta=1$, soft cutoff threshold $z_{\mathrm{cut}}<0.1$, and characteristic radius $R_{0}=0.8$~\cite{Larkoski:2014wba}, which is applied to remove soft, wide-angle radiation from the jet. This is a generalization of the ``modified mass" drop tagger algorithm ~\cite{Dasgupta:2013ihk,Butterworth:2008iy}, and the two are identical when the angular exponent $\beta=0$. This algorithm is based on reclustering the constituents of the AK8 jets using the Cambridge--Aachen algorithm~\cite{Dokshitzer:1997in,Wobisch:1998wt}.
The soft drop jet mass $\mJ$, calculated as the invariant mass of the four-momenta sum of the final remaining jet constituents, weighted according to the PUPPI algorithm, is utilized in the offline analysis and will be denoted as jet mass in the following. The mass is corrected for \PT- and $\eta$-dependent nonuniformities due to detector effects, following the procedure described in Ref.~\cite{CMS-PAS-JME-16-003}.

This algorithm is used for the offline analysis, while the jet-trimming algorithm~\cite{Krohn:2009th} is used at trigger level, see Section~\ref{sec:trigger}. The jet-trimming algorithm reclusters each AK8 jet starting from all its original constituents using the \kt{} algorithm~\cite{Catani:1993hr} to create subjets with a size parameter $R_\text{subjet}$ set to 0.2, discarding any subjet with $\PT^\mathrm{subjet}/\PT^\mathrm{jet}<0.03$.

\begin{figure}[htbp]
\centering
\includegraphics[width=0.49\textwidth]{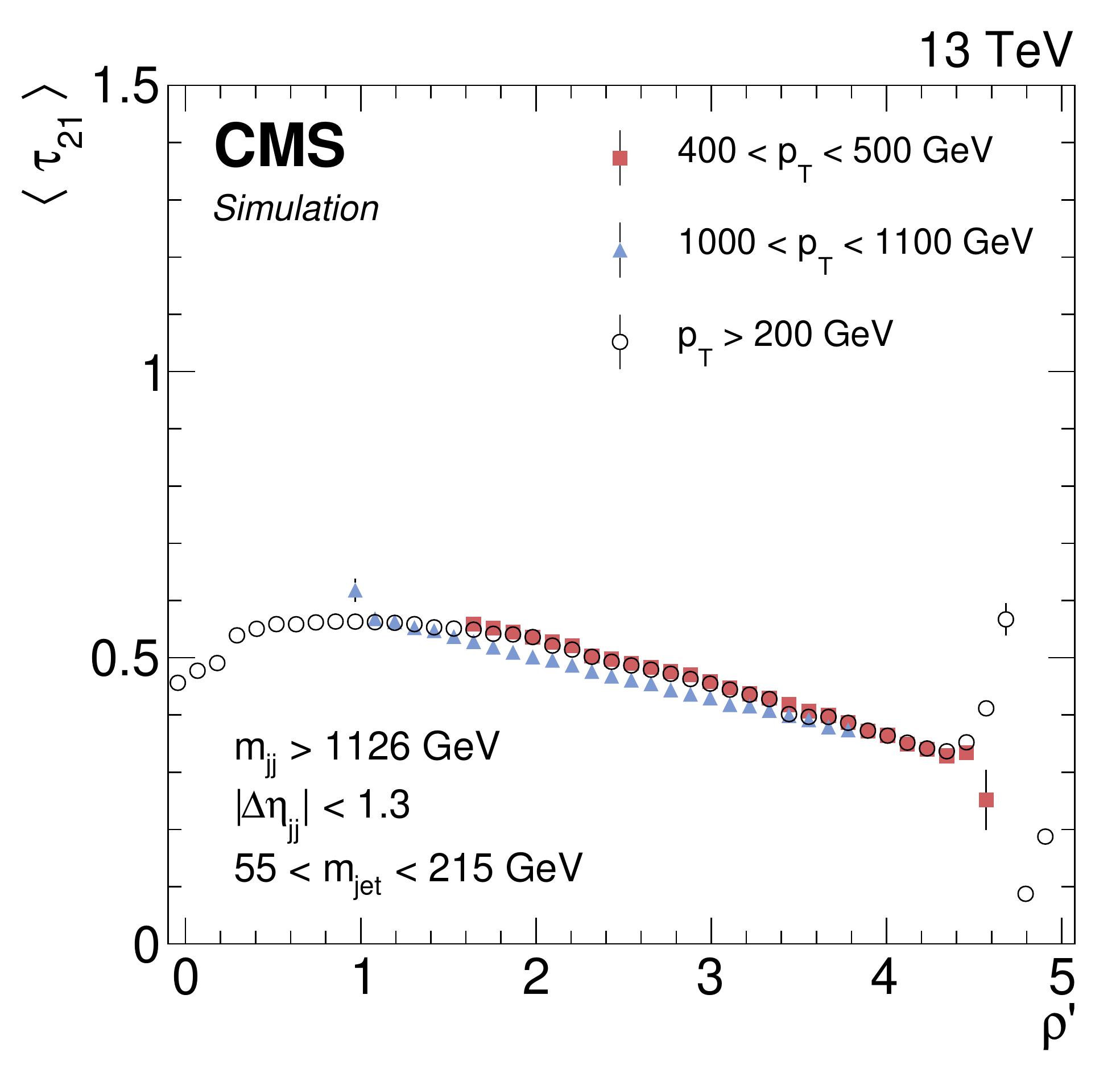}
\includegraphics[width=0.49\textwidth]{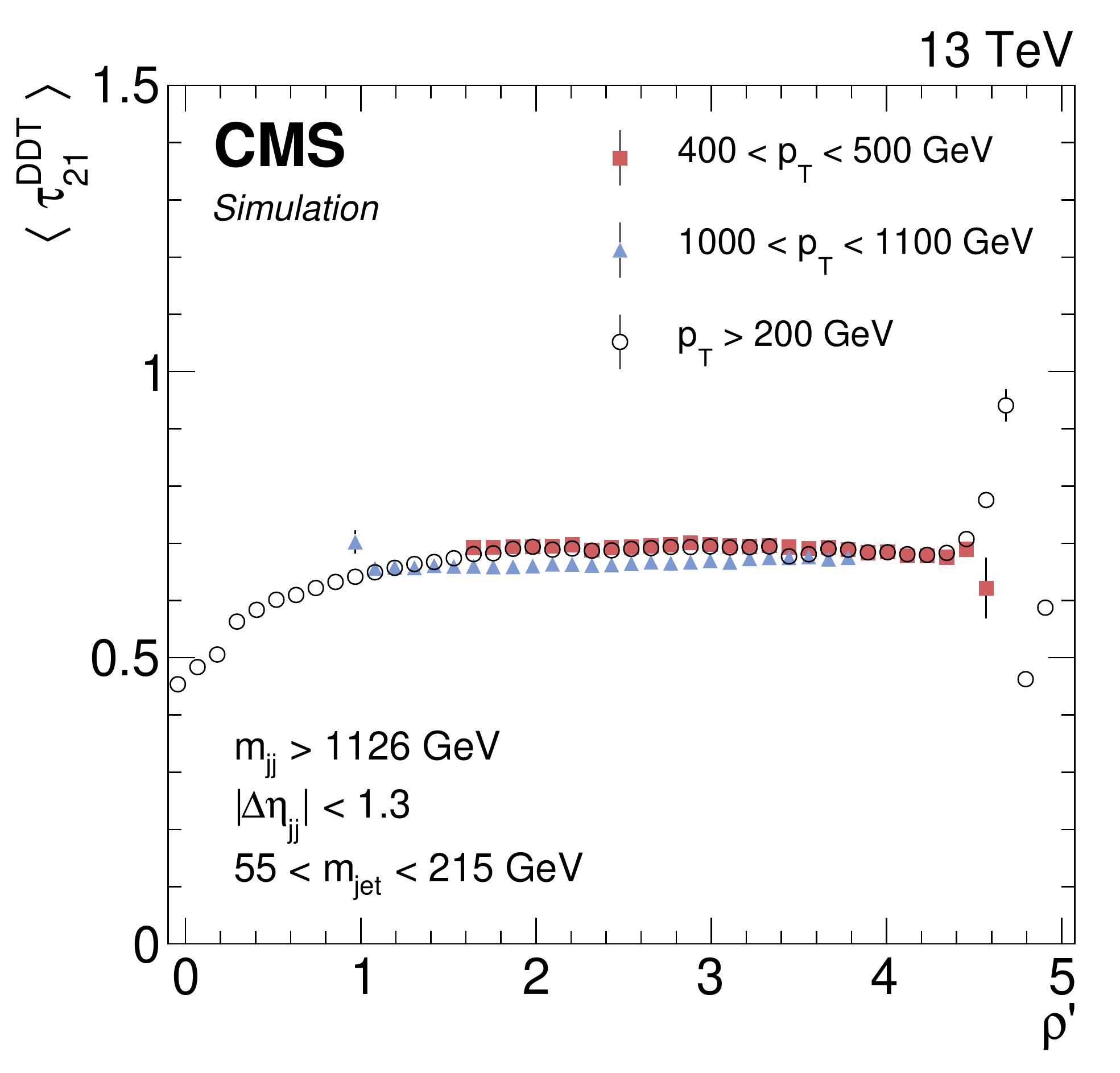}
\caption{The \nsubj (\cmsLeft) and \nsubjDDT (\cmsRight) profile dependencies on $\rho'=\ln(\MJ^2/(\PT\mu))$ examined in QCD multijet events simulated with \PYTHIA{}. A fit to the linear part of the spectrum for $\PT>200\GeV$ yields the slope $M=-0.080$, which is used to define the mass- and \PT-decorrelated variable $\nsubjDDT=\nsubj-M\,\rho'$. }
\label{fig:rho}
\end{figure}

The $N$-subjettiness variable, $\tau_{N}$, is defined as
\begin{equation}
\tau_N=\frac{1}{r_{0}}\sum_{k}\PTk \min(\Delta R_{1,k},\Delta R_{2,k},\ldots,\Delta R_{N,k}),
\end{equation}
where the index $k$ runs over the PF constituents of the jet, and the distances $\Delta R_{n,k}$ are calculated relative to the axis of the $n$-th subjet.
The normalization factor $r_{0}$ is calculated as $r_{0}=R_{0}\sum_{k} \PTk $, setting $R_{0}$ to the distance parameter used in the clustering of the original jet.
The variable $\tau_{N}$ quantifies the compatibility of the jet clustering with the hypothesis that exactly $N$ subjets are present, with small values of $\tau_{N}$ indicating greater compatibility.
The ratio between 2- and 1-subjettiness, $\nsubj=\tau_{2}/\tau_{1}$, is found to be a powerful discriminant between jets originating from \PV decays into quarks (V boson jets) and jets developed from prompt quarks and gluons (quark/gluon jets).
Jets from \PW or \PZ decays in signal events are characterized by lower values of $\nsubj$ relative to SM backgrounds.
However, the $\nsubj$ variable shows a dependence on the jet \PT-scale as well as the jet mass. This particularly affects the monotonically falling behaviour of the nonresonant background distributions. Since this search probes a large range of jet masses and dijet invariant masses ($\MVV$), we decorrelate $\nsubj$ from the jet \PT-scale and jet mass following the ``designed decorrelated taggers (DDT)" methodology presented in Ref.~\cite{ddt}.
We thereby reduce the \nsubj{} profile dependence on $\rho'=\ln(\MJ^2/(\mathrm{p}_T\mu))$, where $\mu=1\GeV$. This leads to the following definition of \nsubjDDT:
\begin{equation}
\label{eq:ddt}
\nsubjDDT =\nsubj - M\,\rho',
\end{equation}
where $M$ is the extracted slope from a fit to the \nsubj{} profile versus $\rho'$ in QCD multijet events simulated with \PYTHIA{} after applying the full analysis selections. It is evaluated to be $M=-0.080$. The \nsubj (\cmsLeft) and \nsubjDDT (\cmsRight) profile dependencies on $\rho'$ are shown in Fig.~\ref{fig:rho}. We observe a small residual difference between intervals of \PT, but this has a negligible impact on the analysis.

\begin{figure}[htbp]
\centering
\includegraphics[width=0.49\textwidth]{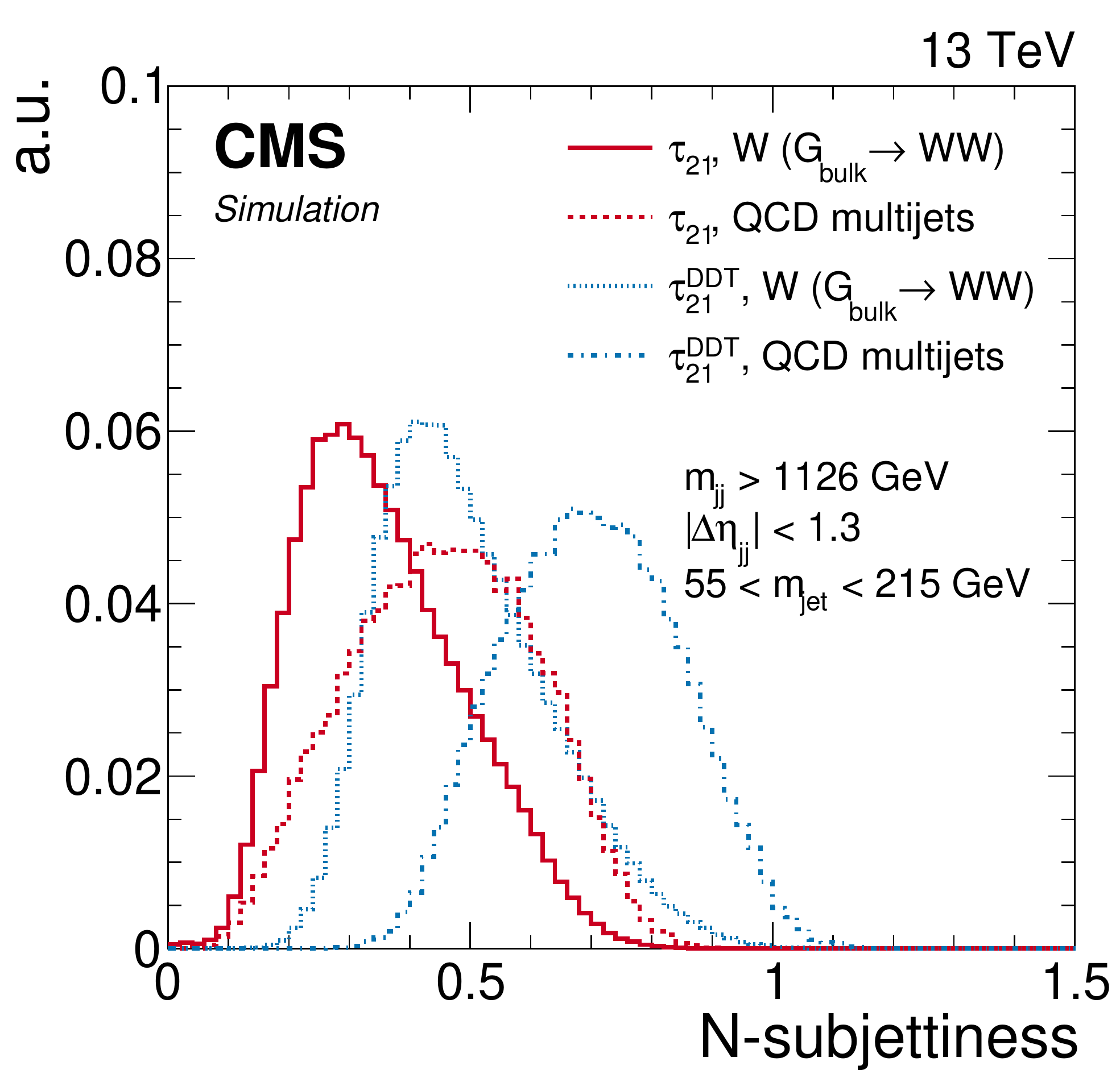}
\includegraphics[width=0.49\textwidth]{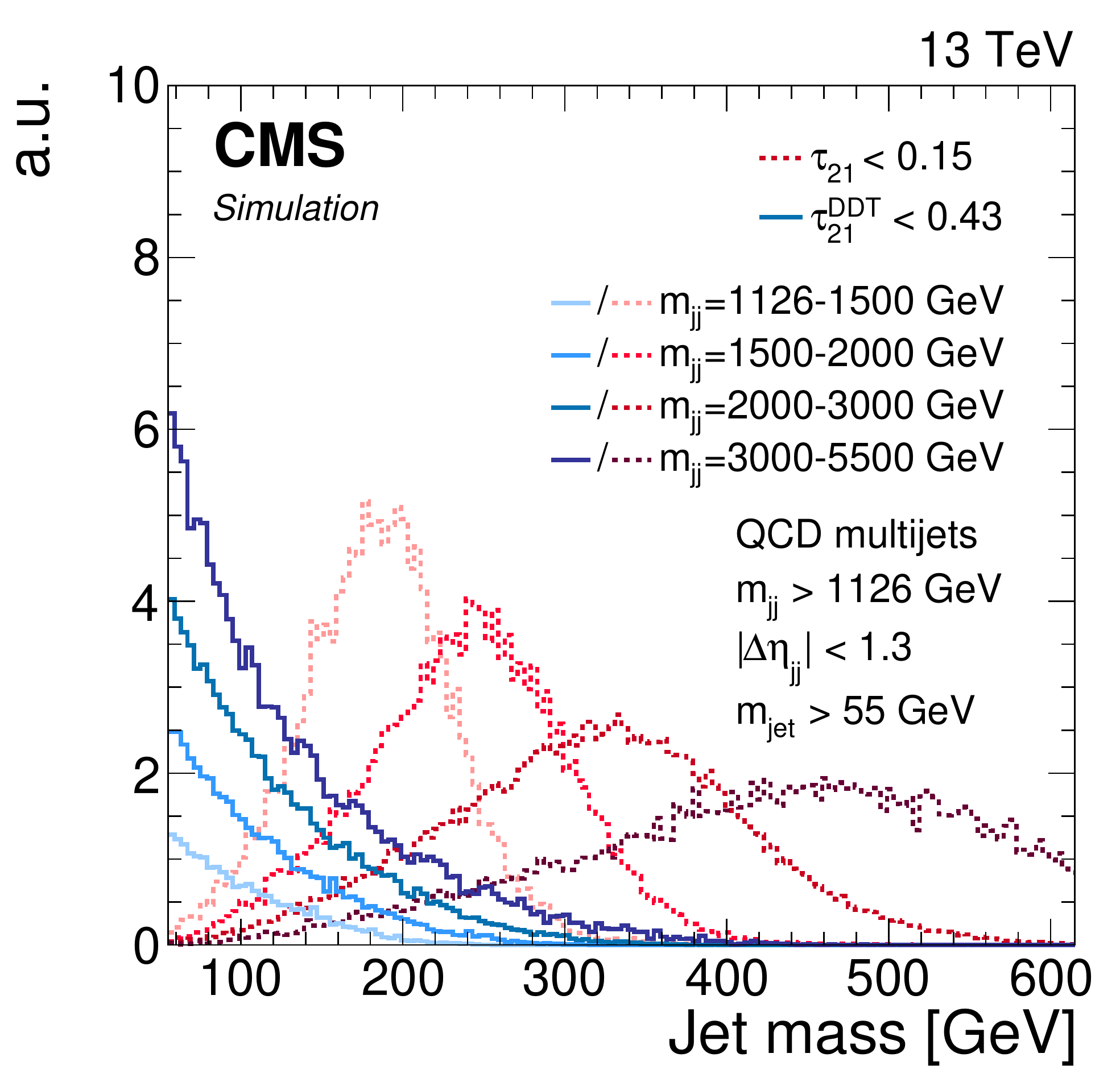}
\caption{Distribution of the $N$-subjettiness discriminants (\nsubj and \nsubjDDT) for \PW-jets and quark/gluon jets from QCD multijet events (\cmsLeft). Distributions in the jet mass of QCD multijet events for four \mjj\ bins in the range 1126--5500\GeV after a cut on \nsubj and \nsubjDDT corresponding to the same mistag rate of about 2\% (\cmsRight). For both discriminants, darker colours correspond to higher \mjj\ ranges. The distributions are arbitrarily scaled for better readability. The analysis selections applied to derive these distributions are specified in the plots. For this analysis the working point of \nsubjDDT${\leq}0.43$ is chosen.}
\label{fig:roc}
\end{figure}

We observe a significant gain in analysis sensitivity when using \nsubjDDT. Since this variable is a function of both $N$-subjettiness and the ratio of jet mass and transverse momentum, it leads to a larger separation between signal and background as shown in the comparison of \nsubjDDT and \nsubj in Fig.~\ref{fig:roc} (\cmsLeft).
Furthermore, using the \nsubjDDT variable reduces the dependency of \mJ\ on \mjj, leading to smoothly falling distributions in the jet mass as shown in Fig.~\ref{fig:roc} (\cmsRight).

\subsection{Trigger and preliminary offline selection}
\label{sec:trigger}

Events are selected online with a variety of different jet triggers based on the highest jet \PT or the \PT sum of all jets in the event (\HT{}). For some of these triggers additional requirements on the trimmed mass are applied in order to be able to lower the \PT and \HT thresholds. For example, for 2017 data taking, requiring the trimmed jet mass of the leading-\PT jet to be above 30\GeV allows the lowering of the \PT threshold from 500 to 360\GeV while maintaining a similar rate. In the case of the \HT{}-triggers, the threshold can be lowered from 1050 to 750--800\GeV when requiring a trimmed jet mass greater 50\GeV. The \HT{}-triggers utilize a standard jet collection of anti-\kt jets with a distance parameter $R = 0.4$, while the triggers based on the trimmed jet mass operate on AK8 jets.
The triggers used for the 2017 data set are conceptually similar to those used in 2016, and correspond to those used in Ref.~\cite{Sirunyan:2017acf}. The 2017 trigger thresholds were slightly greater than those in 2016 in order to maintain the same trigger rate despite a higher instantaneous luminosity.

\begin{figure}[htbp]
\centering
\includegraphics[width=0.49\textwidth]{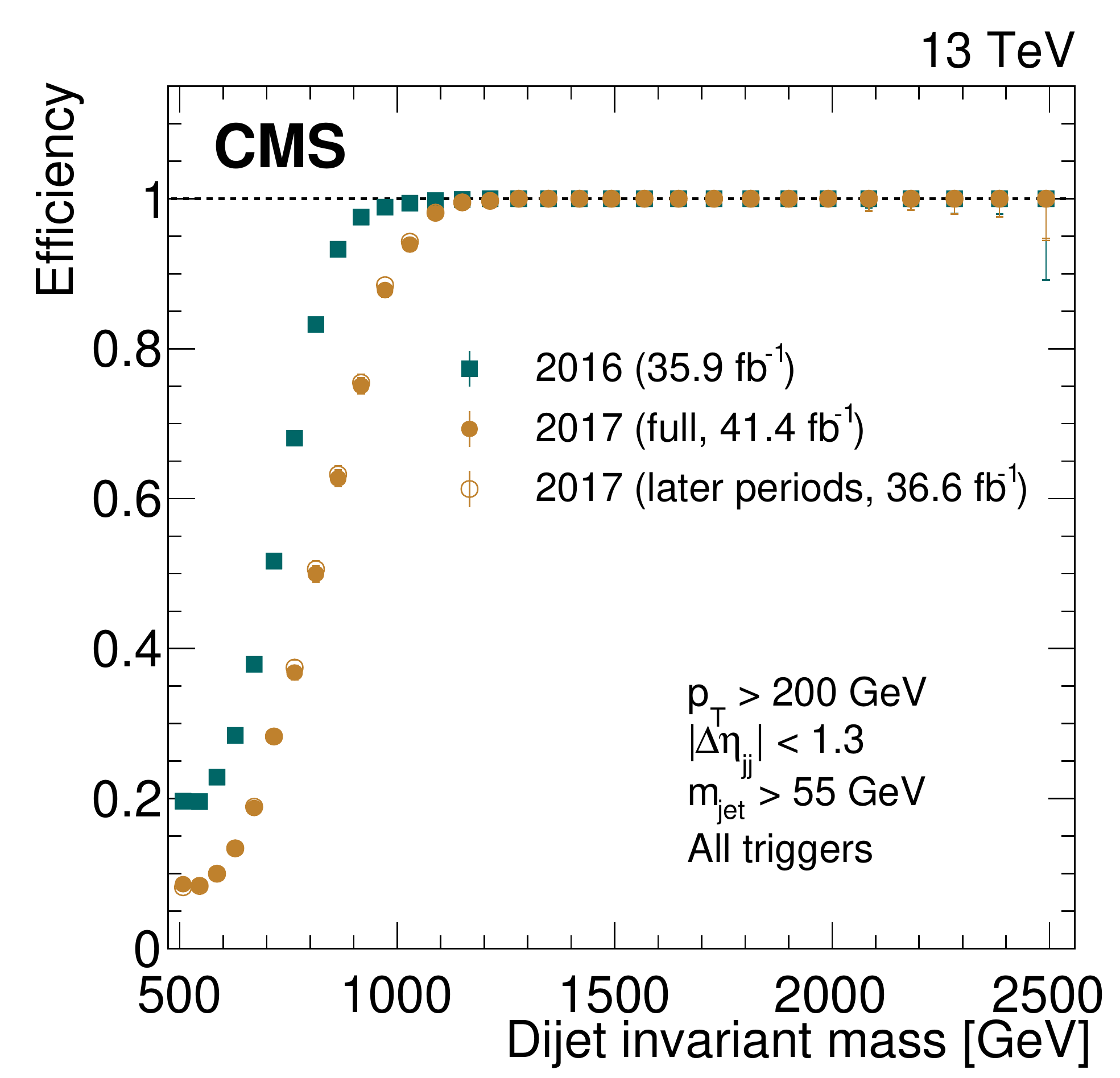}
\includegraphics[width=0.49\textwidth]{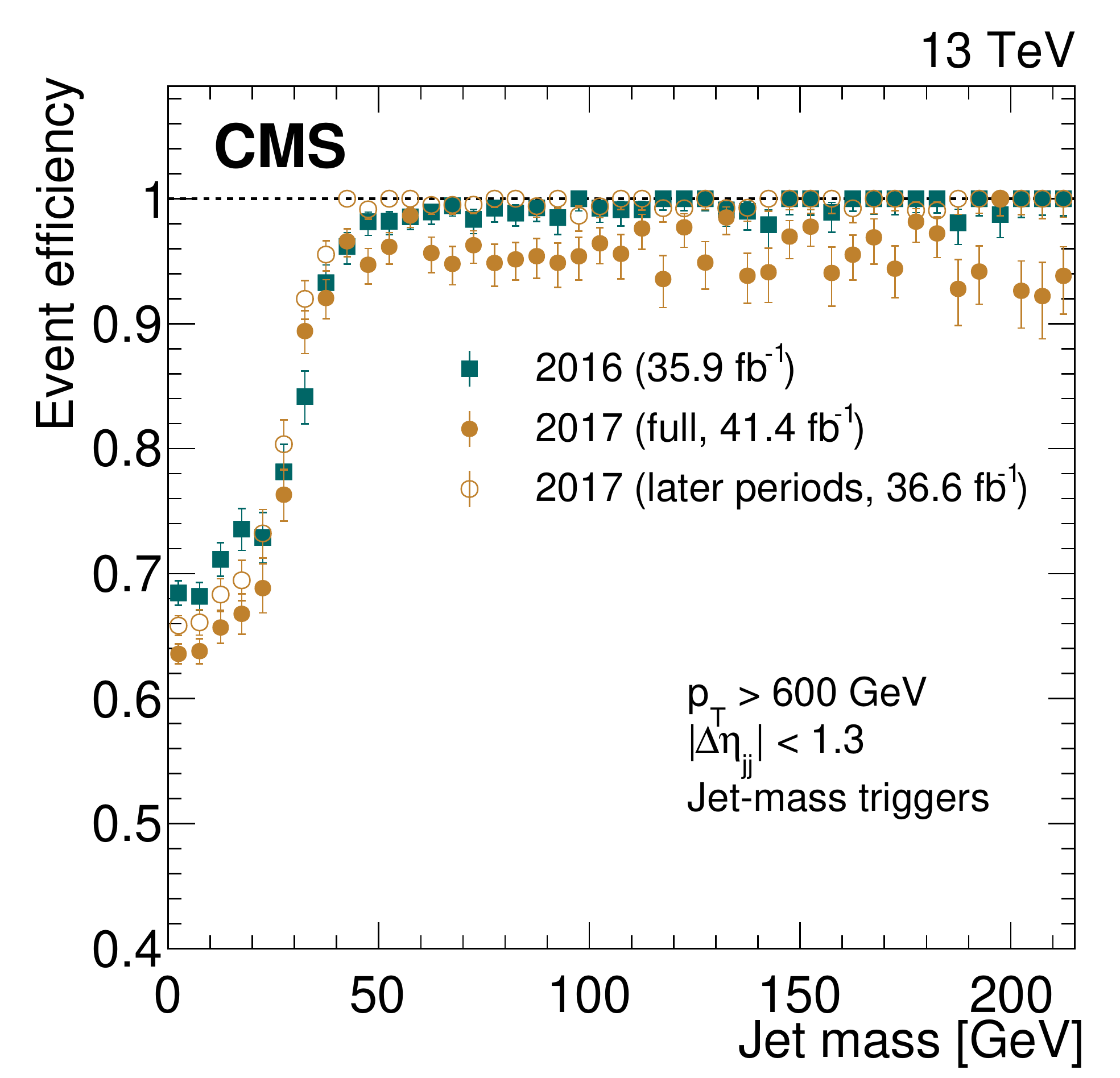}
\caption{The trigger efficiency as a function of the dijet invariant mass for a combination of all triggers used in this analysis (\cmsLeft) and the event efficiency for either of the selected jets to pass triggers requiring an online trimmed mass of at least 30\GeV as a function of the jet mass (\cmsRight). The solid yellow circles correspond to the trigger efficiencies for the full 2017 data set and do not reach 100\% efficiency because the jet mass based triggers were unavailable for a period at the beginning of data taking (corresponding to 4.8\fbinv). The open yellow circles are the corresponding efficiencies excluding this period. The uncertainties shown are statistical only.}
\label{fig:trigturnon}
\end{figure}

The trigger efficiency as a function of the dijet invariant mass is measured in an orthogonal single muon data set, shown in Fig.~\ref{fig:trigturnon}, using a combination of all triggers (\cmsLeft), and as a function of the jet mass for the triggers exploiting the trimmed jet mass (\cmsRight). For the trimmed jet mass triggers, the efficiency plateau as a function of the jet mass does not reach 100\% for the full 2017 data set (full yellow circles), since these triggers were not used for the first 4.8\fbinv of data recorded. The trigger efficiency excluding this period is shown with open yellow circles. The combination of all triggers is $>99$\% efficient above dijet invariant masses of 990 and 1126\GeV for the full 2016 and 2017 data sets, respectively. For simplicity, the subsequent analysis requires the dijet invariant mass to be above 1126\GeV for both data sets. Given the \mjj resolution of about 10\%, the lowest resonance mass that is accepted with high efficiency by the analysis is 1.2\TeV.

All events are required to have at least one primary vertex reconstructed within a 24\cm window along the beam axis, with a transverse distance from the average \Pp{}\Pp interaction region of less than 2\cm~\cite{Chatrchyan:2014fea}. The reconstructed vertex with the largest value of summed physics-object $\PT^2$ is taken to be the primary \Pp{}\Pp interaction vertex. The physics objects are the jets, clustered using the jet finding algorithm~\cite{Cacciari:2008gp,Cacciari:2011ma} with the tracks assigned to the vertex as inputs, and the associated missing transverse momentum, taken as the negative vector sum of the \pt of those jets.

\subsection{Event selection}

\begin{figure*}[htbp]
\centering
\includegraphics[width=0.49\textwidth]{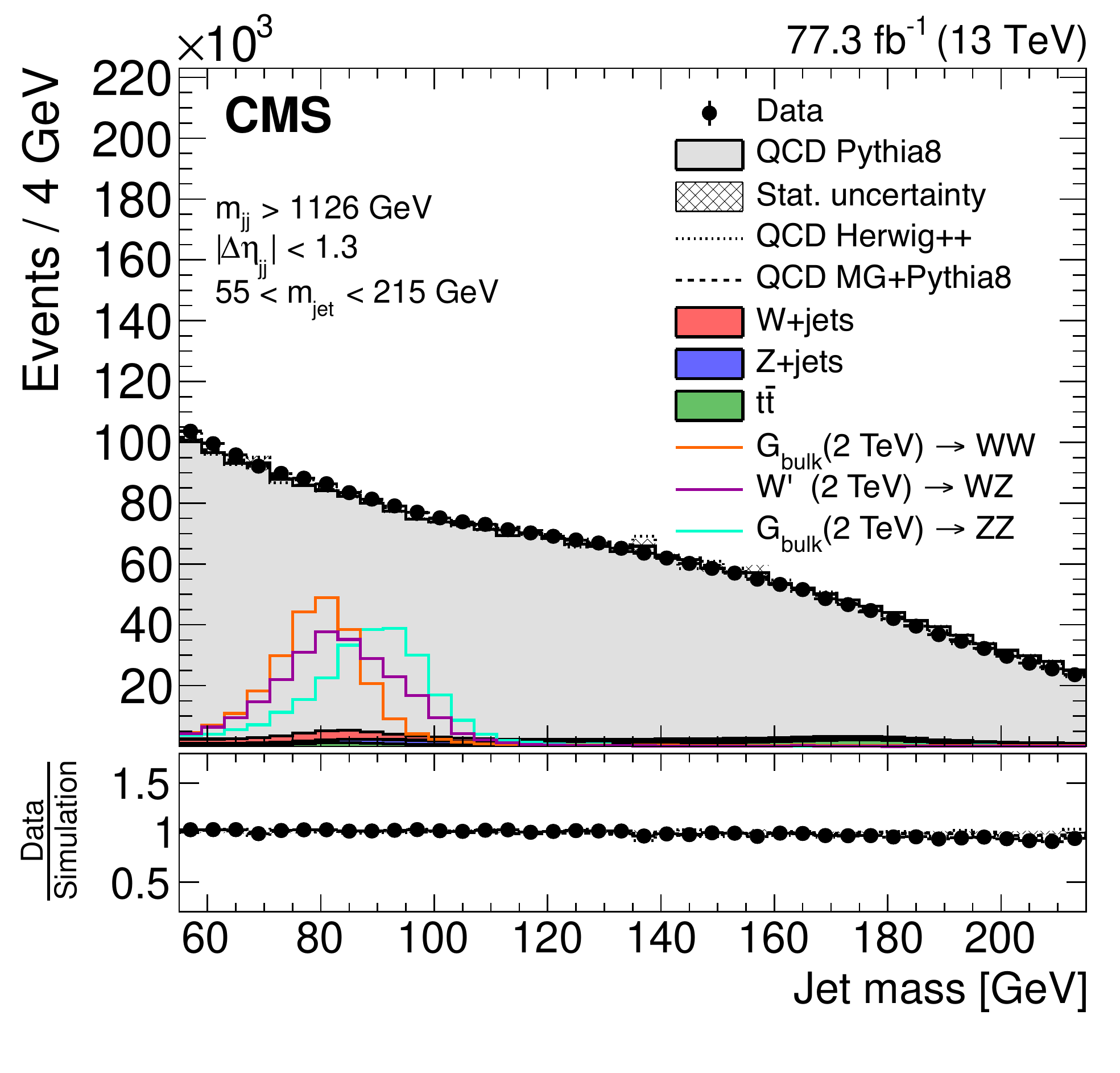}
\includegraphics[width=0.49\textwidth]{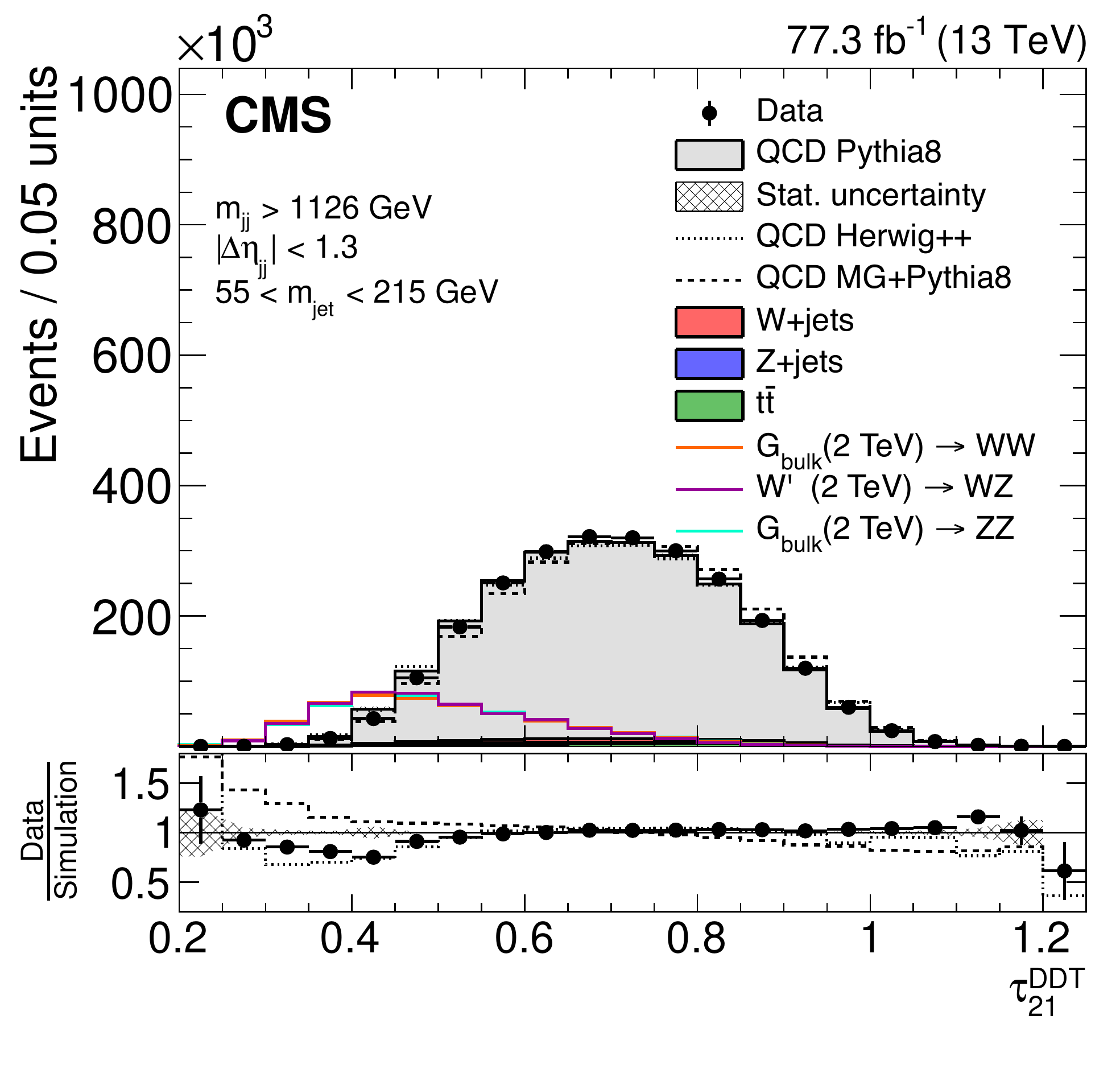}\\
\includegraphics[width=0.49\textwidth]{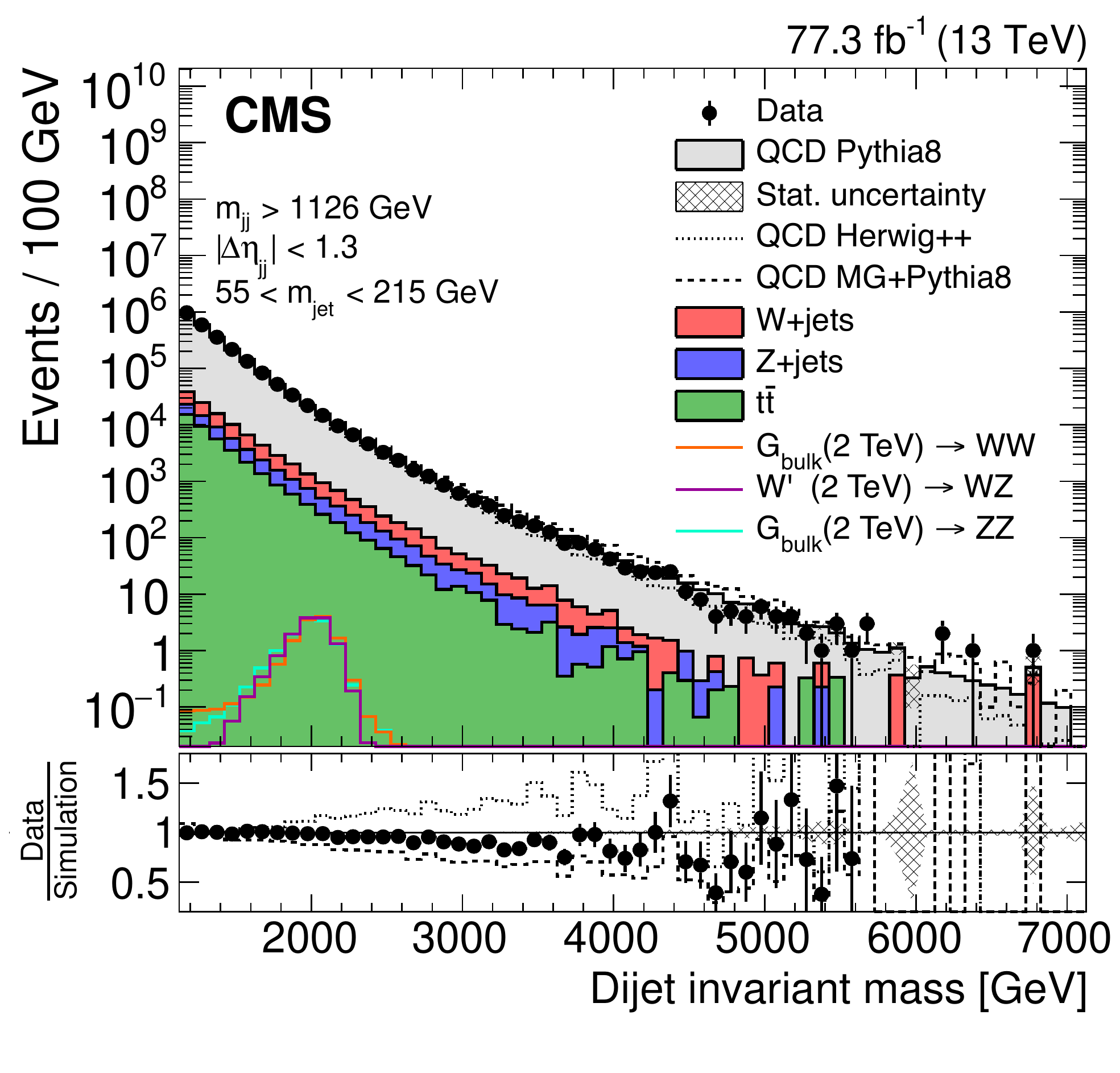}
\caption{Jet mass (upper left) and \nsubjDDT (upper right) distributions for selected jets (one random jet per event), and dijet invariant mass distribution (lower), for events with a jet mass between 55 and 215\GeV. For the QCD multijet simulation, several alternative predictions are shown, scaled to the data minus the other background processes, which are scaled to their SM expectation as described in the text. The different signal distributions are scaled to be visible. No selection on \nsubjDDT is applied. The ratio plots show the fraction of data over QCD multijet simulation for \PYTHIA{8} (black markers), \HERWIG{++} (dotted line), and \MADGRAPH{}+\PYTHIA{8} (dashed line).}
\label{fig:nsubj}
\end{figure*}

Events are selected by requiring at least two jets with $\PT>200\GeV$ and $\abs{\eta}<2.5$. The two jets with the highest \PT in the event are selected as potential vector boson candidates and are required to have a separation of $\abs{\Delta\eta}<1.3$ in order to reduce the QCD multijet background. In addition to the requirement that the two jets invariant mass $\mjj>1126\GeV$, based on the trigger selection discussed above, it is further required that $\mjj<5500\GeV$. The upper cut on mjj is well above the highest dijet mass event observed in data. To simplify the modelling of the 3D shapes in the $\MVV$-$\MJO$-$\MJT$ space, the two jets are labelled at random so that the mass distributions of the first and second selected jet, \MJO and \MJT, have the same shape.
Jets originating from the misreconstruction of a high momentum lepton are rejected by requiring an angular separation $\Delta R>0.8$ to muons (electrons) with $\PT$ greater than 20 (35)\GeV and satisfying identification criteria optimized for high-momentum leptons~\cite{Sirunyan:2018fpa,Khachatryan:2015hwa}.
To reduce the QCD multijet background, we require the jet mass to be between 55 and 215\GeV. The selected events are further grouped into two categories according to their likelihood to originate from a boson decay into quarks, as quantified by \nsubjDDT{}. The jet mass, \nsubjDDT, and dijet invariant mass distributions for data and simulation are shown in Fig.~\ref{fig:nsubj}.

In the high-purity (HPHP) category, both jets are required to have $0<\nsubjDDT\leq0.43$, while in the low-purity (HPLP) category only one of the jets needs to fulfill this requirement, and the other must satisfy $0.43<\nsubjDDT\leq0.79$. These conditions yield the highest expected signal significance over the whole mass range, while at the same time selecting at least 95\% of the signal. The addition of the HPLP category improves the expected cross section upper limit by around 20\% at high dijet invariant mass where the background is low. Finally, a loose requirement of $\rho=\ln(\MJ^2/\PT^2) < -1.8$ is applied in order to veto events in which the jet mass is high, but the jet \PT is low. In these cases the cone size of $\Delta R=0.8$ is too small to contain the full jet, affecting both the jet mass resolution and the \nsubjDDT tagging efficiency, which is not well modelled in simulation. This selection has a negligible effect on the signal, which typically has jets with masses around the \PW or \PZ boson mass and high \PT{}.

\subsection{Substructure variable corrections and validation}
\label{subsec:SubVal}

Figure~\ref{fig:nsubj} shows a notable deviation in the shape of the \nsubjDDT distribution between data and simulation. Such mismodelling introduces a bias in the jet tagging efficiency for the signal, and as a consequence in the measured signal rate. We therefore compute scale factors to correct the signal efficiency accordingly. For the background jets, this mismodelling requires no further correction, because of the data-driven approach adopted in this analysis, where the background shape and normalization are fitted to data with large pre-fit uncertainties as described in the following sections.

The \PW\ boson tagging scale factors and jet mass scale and resolution uncertainties are estimated from data by isolating a control sample of merged \PW bosons in a high-\PT{} \ttbar sample. This is done by performing a simultaneous fit to the jet mass distributions for the two ranges of \nsubjDDT, as detailed in Ref.~\cite{CMS-PAS-JME-16-003}.
To extract the efficiency from a clean sample of merged W bosons, the \ttbar sample is split into two components, depending on whether the quarks from the W boson decay at truth level are within $\delta R = 0.8$ or not, \ie on whether the hadronic boson decay is merged into a single jet or not. Only the merged component is considered in the efficiency calculation and the mass scale and resolution extraction.

\begin{figure*}[htbp]
 \centering
 \begin{tabular}{cc}
\includegraphics[width=0.465\textwidth]{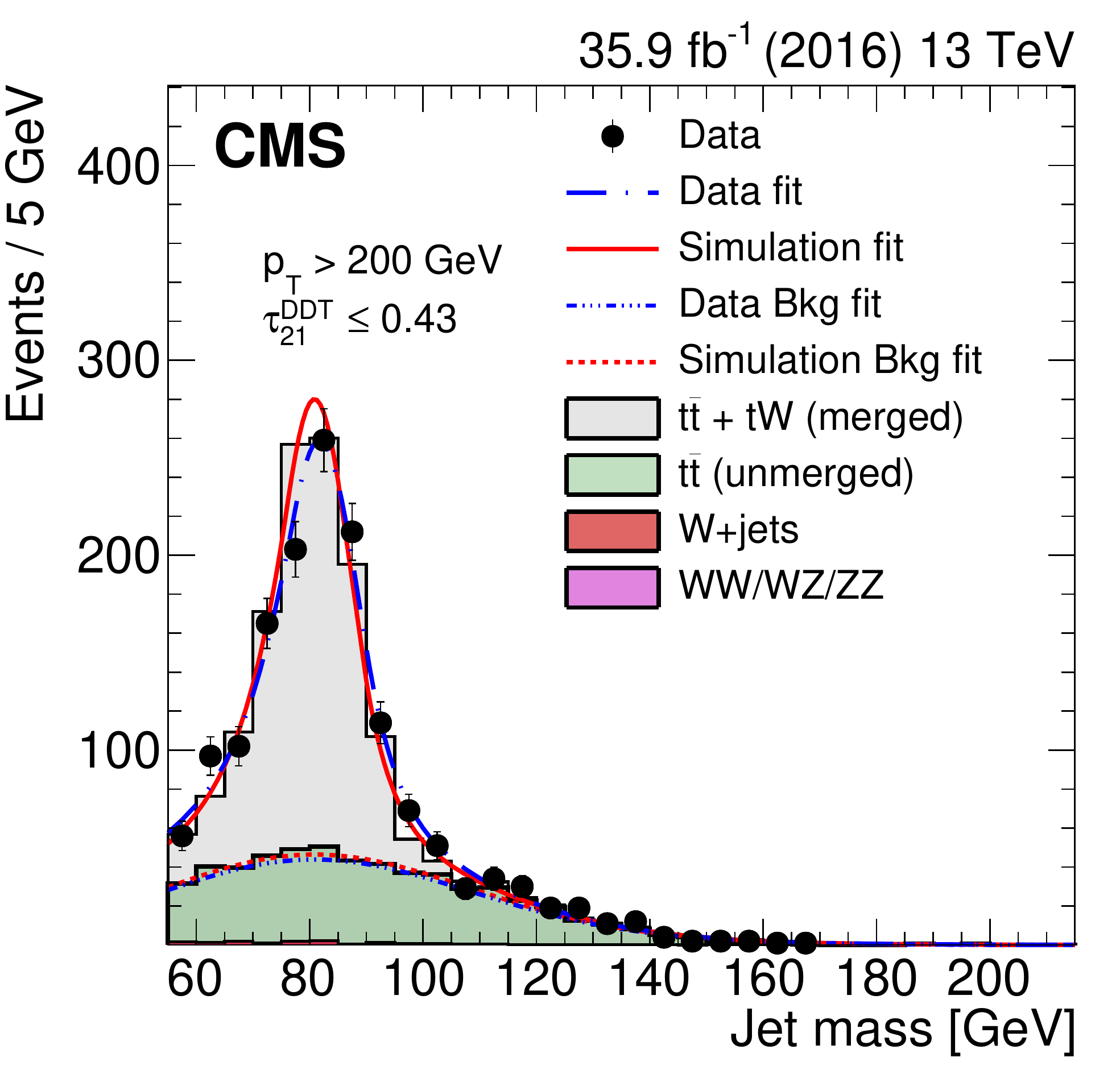}
\includegraphics[width=0.49\textwidth]{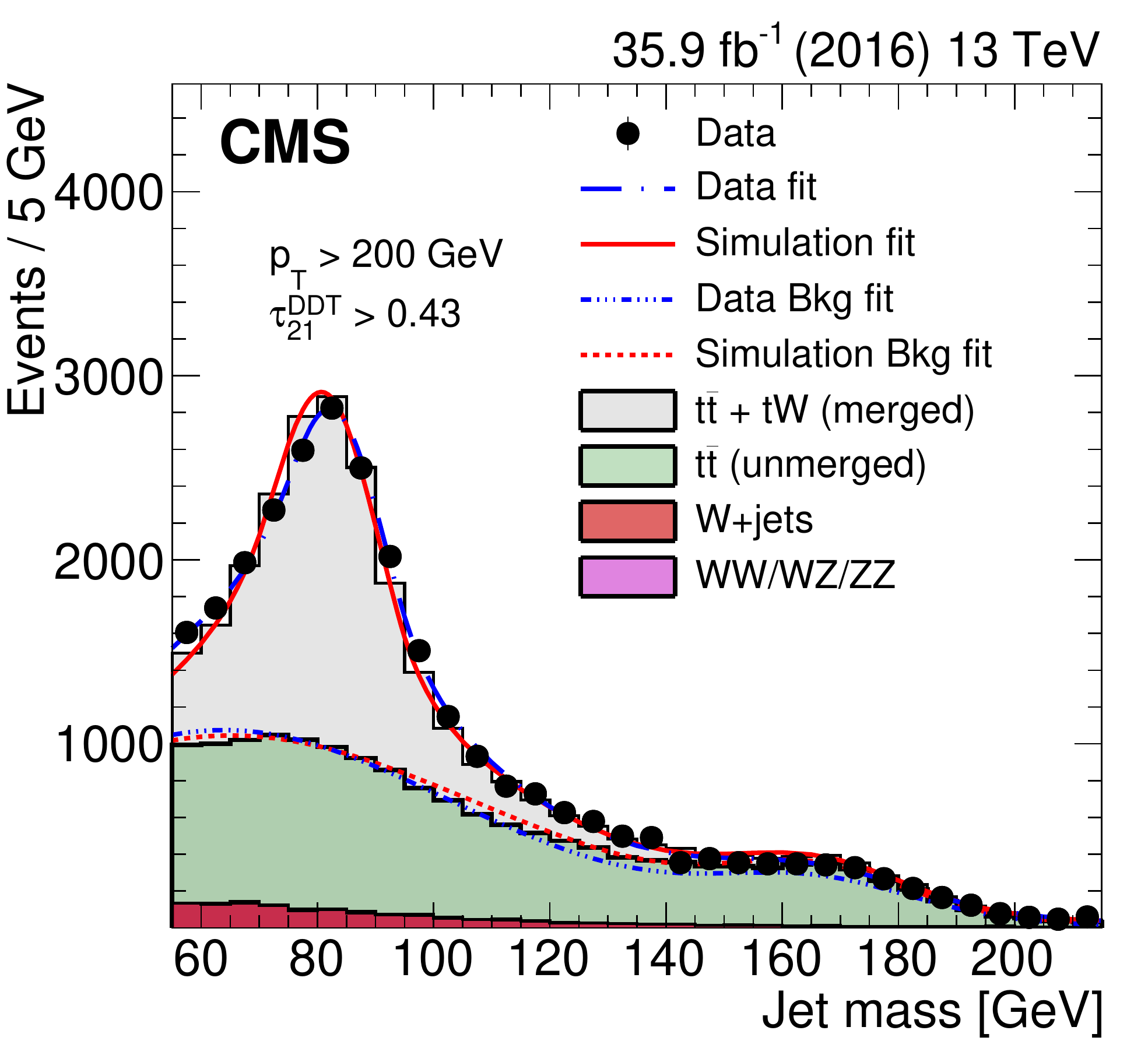}\\
\includegraphics[width=0.465\textwidth]{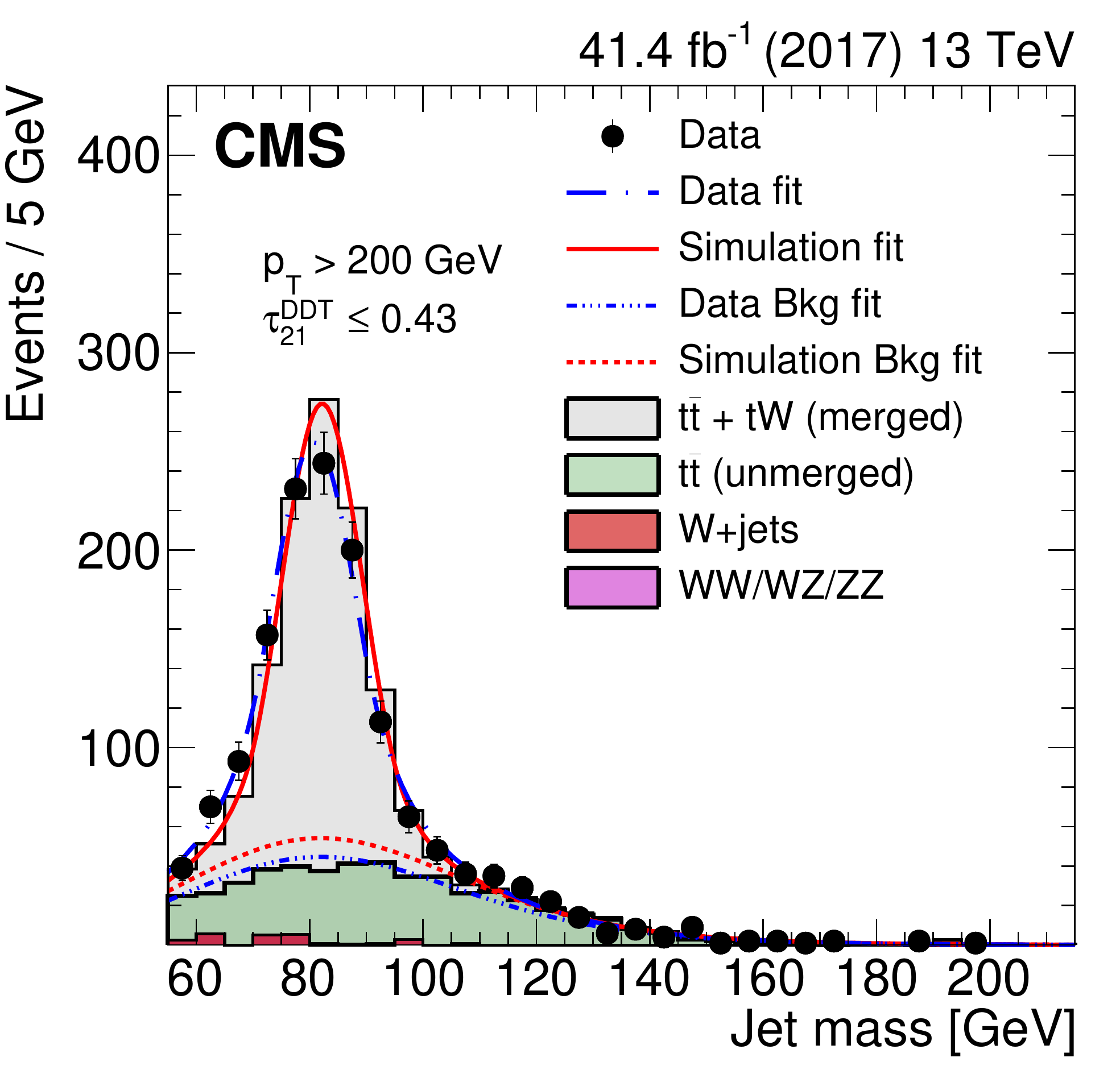}
\includegraphics[width=0.49\textwidth]{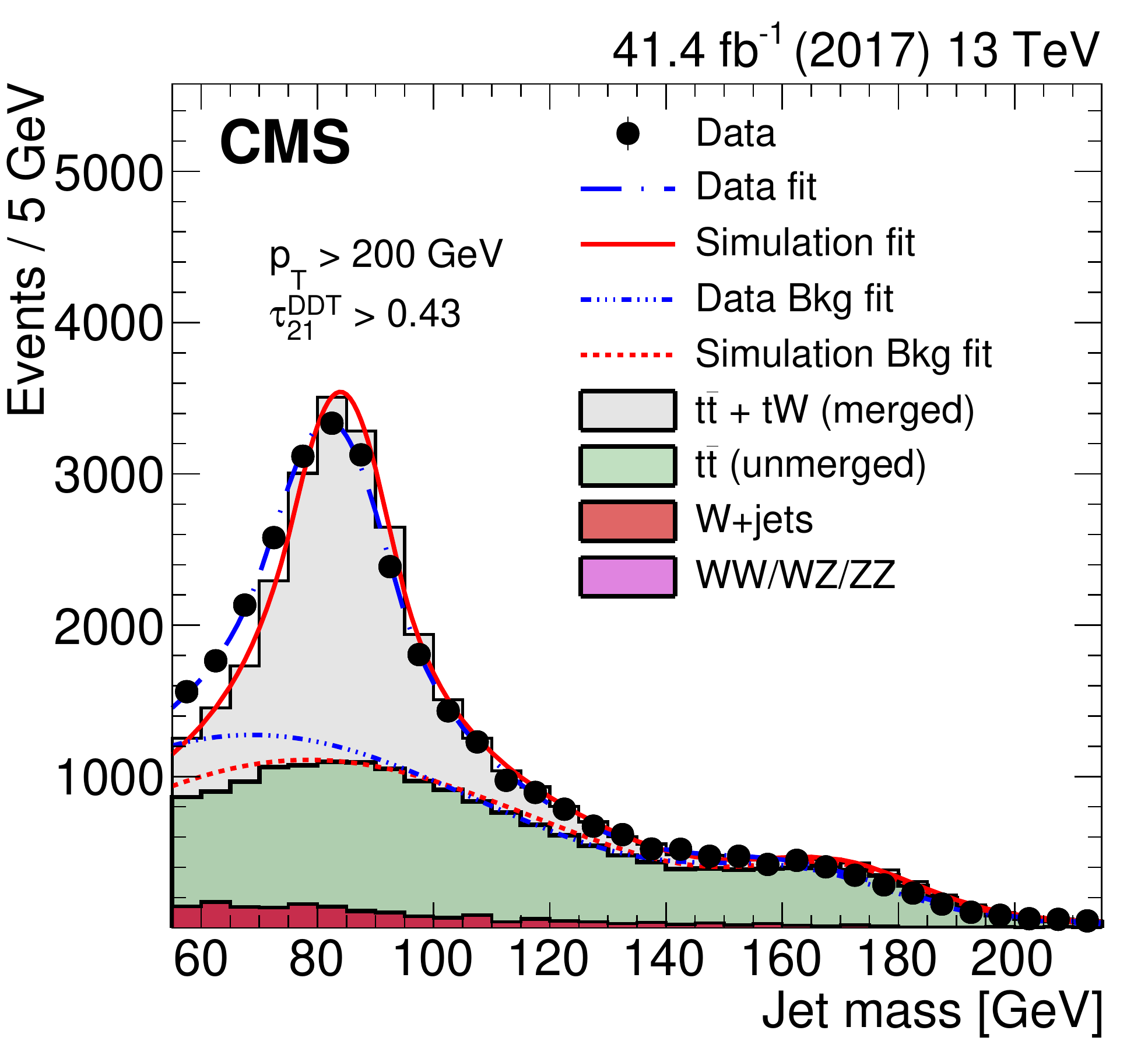}
 \end{tabular}
 \caption{The jet mass distribution for events that pass (left) and fail (right) the $\nsubjDDT\leq0.43$ selection in the \ttbar control sample. The results of the separate fits to data and to simulation are shown by the dash-dotted blue and solid red lines, respectively. The background components of the fits are shown as dashed and dash-dotted lines. The fit to 2016 data is shown in the upper panels and the fit to 2017 data in the lower panels. The associated uncertainties are shown in Table~\ref{tab:wsf_total} and discussed further in Section~\ref{subsec:SubVal}.}
 \label{fig:simFit}
 \end{figure*}

The efficiencies and scale factors obtained are listed in Table~\ref{tab:wsf_total} for 2016 and 2017 data, with the corresponding fits shown in Fig.~\ref{fig:simFit}. The \PW\ boson tagging efficiency in the selected \ttbar events of around 7\% is relatively low compared to the efficiency in signal events, since these events are dominated by \PW\ boson jets with a \PT of around 200\GeV, just at the threshold where the decay products of the \PW\ boson merge into a single jet. The signal jets, however, mostly have a \PT above 600\GeV, and a tagging efficiency around 35\%. The signal efficiency for \nsubjDDT increases with the jet \PT{}, whereas the background efficiency is constant, as shown in Ref.~\cite{CMS-PAS-JME-16-003}. Two systematic uncertainties in the scale factors are added: one due to differences in MC generation and modelling of the parton shower and one due to NNLO corrections. The former is evaluated by comparing the resulting scale factors when using \ttbar simulation produced with different generators. The latter is evaluated by comparing the extracted efficiencies with and without reweighting according to the top quark \PT, where the reweighting is derived from data in order to better describe the observed \PT distribution in \ttbar data~\cite{Khachatryan:2016mnb}.
The jet mass scale and resolution are estimated in the same fits and also listed in Table~\ref{tab:wsf_total}. The difference in jet mass scale between data and simulation is around 2\%, and the jet mass resolution difference is roughly 8\%. These are used to scale and smear the jet mass in simulation, and their uncertainties are additionally inserted as systematic uncertainties in the final fit.

\begin{table*}[htbp]
  \centering
  \topcaption{The \PW jet mass peak position (m) and resolution ($\sigma$), and the \PW{}-tagging efficiencies, as extracted from top quark enriched data and from simulation, together with the corresponding data-to-simulation scale factors. The uncertainties in the scale factors include systematic uncertainties estimated as described in Ref.~\cite{CMS-PAS-JME-16-003}.}
  \cmsTable{
  \begin{tabular}{lccc}
    \hline
    & m [\GeVns{}]           & $\sigma$ [\GeVns{}]     & \PW-tagging efficiency\\
    \multicolumn{4}{c}{2016}\\
    $\nsubjDDT<0.43$ & &&  \rule{0pt}{2.6ex} \rule[-1.2ex]{0pt}{0pt}\\
    Data            & $82.0\pm0.5\stat$   & $7.1\pm0.5\stat$ & $0.080\pm0.008\stat$\\
    Simulation      & $80.9\pm0.2\stat$   & $6.6\pm0.2\stat$ & $0.085\pm0.003\stat$\\
    Data/simulation & $1.014\pm0.007$\,(stat+syst)       & $1.09\pm0.09$\,(stat+syst)     & $0.94\pm0.10$\,(stat+syst)\\ [\cmsTabSkip]
    $0.43<\nsubjDDT<0.79$ & && \rule{0pt}{2.6ex} \rule[-1.2ex]{0pt}{0pt} \\
    Data            &    &  & $0.920\pm0.008\stat$\\
    Simulation      &    &  & $0.915\pm0.003\stat$\\
    Data/simulation &    &   & $1.006\pm0.009$\,(stat+syst)\\
    [\cmsTabSkip]
    \multicolumn{4}{c}{2017}\\
    $\nsubjDDT<0.43$ & &&  \rule{0pt}{2.6ex} \rule[-1.2ex]{0pt}{0pt} \\
    Data            & $80.8\pm0.4\stat$   & $7.7\pm0.4\stat$ & $0.065\pm0.006\stat$\\
    Simulation      & $82.2\pm0.3\stat$   & $7.1\pm0.3\stat$ & $0.068\pm0.005\stat$\\
    Data/simulation & $0.983\pm0.007$\,(stat+syst)       & $1.08\pm0.08$\,(stat+syst)     & $0.96\pm0.12$\,(stat+syst)\\ [\cmsTabSkip]
    $0.43<\nsubjDDT<0.79$ & && \rule{0pt}{2.6ex} \rule[-1.2ex]{0pt}{0pt} \\
    Data            &    &  & $0.935\pm0.006\stat$\\
    Simulation      &    &  & $0.932\pm0.005\stat$\\
    Data/simulation &    &   & $1.003\pm0.008$\,(stat+syst)\\
    \hline
  \end{tabular}
  }
  \label{tab:wsf_total}
\end{table*}

\section{The multi-dimensional fit}
\label{sec:multidimfit}

The background estimation technique used in previous versions of this analysis~\cite{Khachatryan:2014hpa,Sirunyan:2016cao,Sirunyan:2017acf}
relied on a one-dimensional (1D) fit of the dijet invariant mass after a tight jet mass selection (65--105\GeV) has been applied.
In the analysis presented here, we take advantage of the fact that the signal peaks in three observables (the jet masses $\MJO$ and $\MJT$, and the dijet invariant mass $\MVV$),
and attempt to extract the signal by searching for peaks in the multi-dimensional $\MVV$-$\MJO$-$\MJT$ space.
This method permits searches for generic resonances, decaying to two SM or non-SM bosons, anywhere in the jet mass and dijet invariant mass spectra in the future.
Additionally, tight jet mass cuts as used in previous diboson resonance searches are no longer needed, as we fit the full jet mass line shape to extract the signal.
Since such a cut around the vector boson mass leads to about 20\% inefficiency for the \PW\ and \PZ\ boson signals, including all the events that would fall outside the mass window reduces the statistical uncertainties in the fitting procedure. Furthermore, the background \mjj{} shape is better constrained at high dijet invariant masses than it is in the previous method.

Fitting the jet mass and resonance mass together also allows us to add nuisance parameters that simultaneously affect the jet masses and the resonance mass, accounting for their correlation. We build a three dimensional background model starting from simulation. As the number of simulated events is small, a forward-folding kernel approach is used to ensure a full and smooth model, as described in Section~\ref{subsec:QCDtemplates}. Further, to account for discrepancies in the QCD multijet background simulation and data, we allow the background model to adapt to the data using physically motivated shape variations.

The random jet labelling adapted in the analysis results in essentially the same jet mass distributions for jet-1 and jet-2 in the modelling and removes any correlations between the two jet masses. Thus only the distribution for one of the jet masses are shown in the following Figures.

\subsection{Signal modelling}
\label{sec:signal}

\begin{figure}[htbp]
\centering
\includegraphics[width=0.49\textwidth]{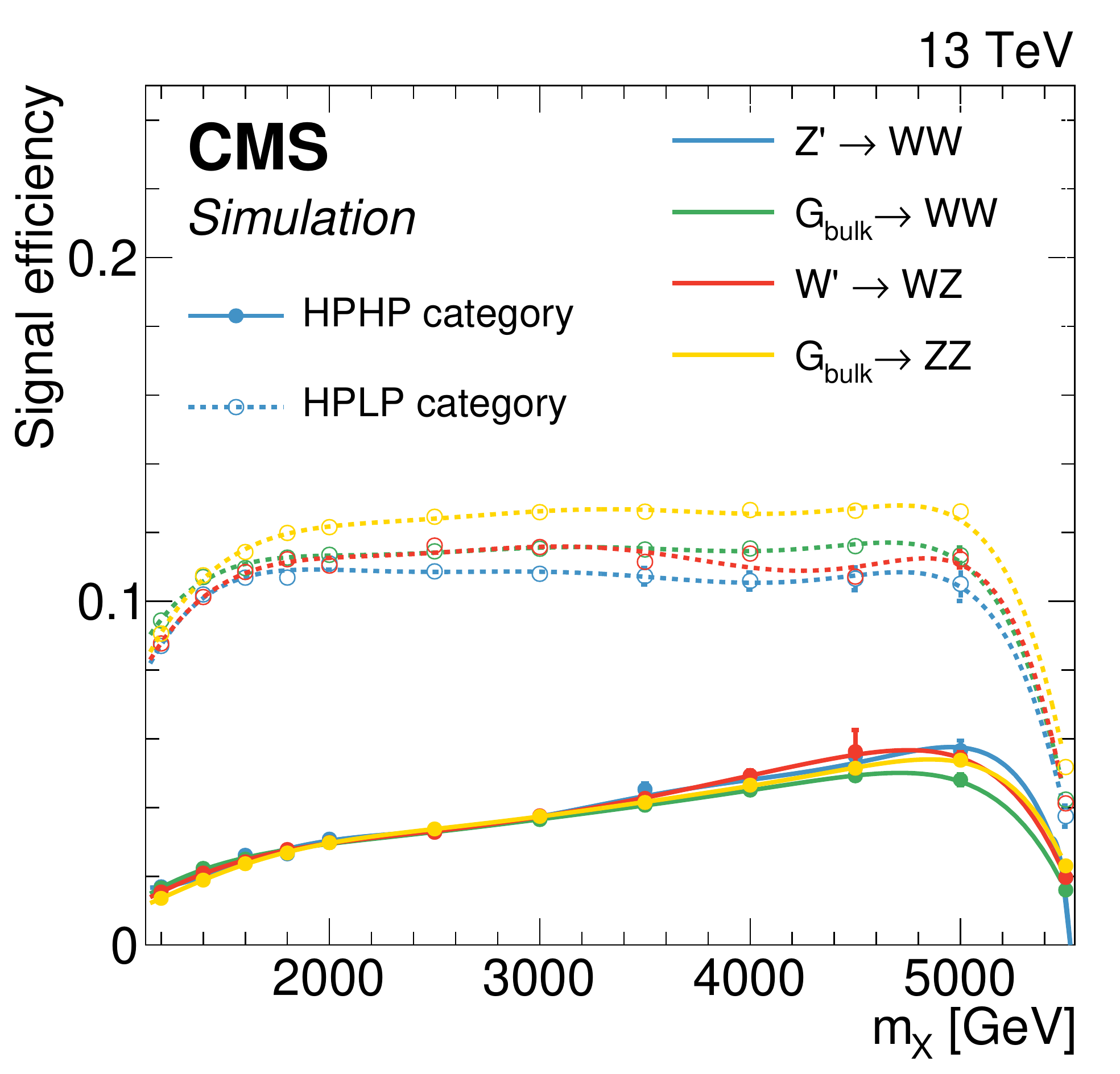}
\caption{Total signal efficiency as a function of \MX after all selections are applied, for signal models with a \PZpr decaying to \PW{}\PW{}, $\BulkG$ decaying to \PW{}\PW{}, $\PWpr$ decaying to \PW{}\Zo{}, and $\BulkG$ decaying to \PZ{}\PZ{}. The denominator is the number of generated events. The solid and dashed lines show the signal efficiencies for the HPHP and HPLP categories, respectively. The decrease in efficiency between 5.0 and 5.5\TeV is due to the requirement $\MVV<5500$\GeV.}
\label{fig:SignalEfficiency}
\end{figure}

For each mass point \MX and each purity category, the signal yield per pb of cross section is calculated as the integral of the histogram produced from the parameterization.
The total signal yield for events passing all analysis selections divided by the number of generated events as a function of \MX is shown in Fig.~\ref{fig:SignalEfficiency}.

\begin{figure}[htbp]
\centering
\includegraphics[width=0.49\textwidth]{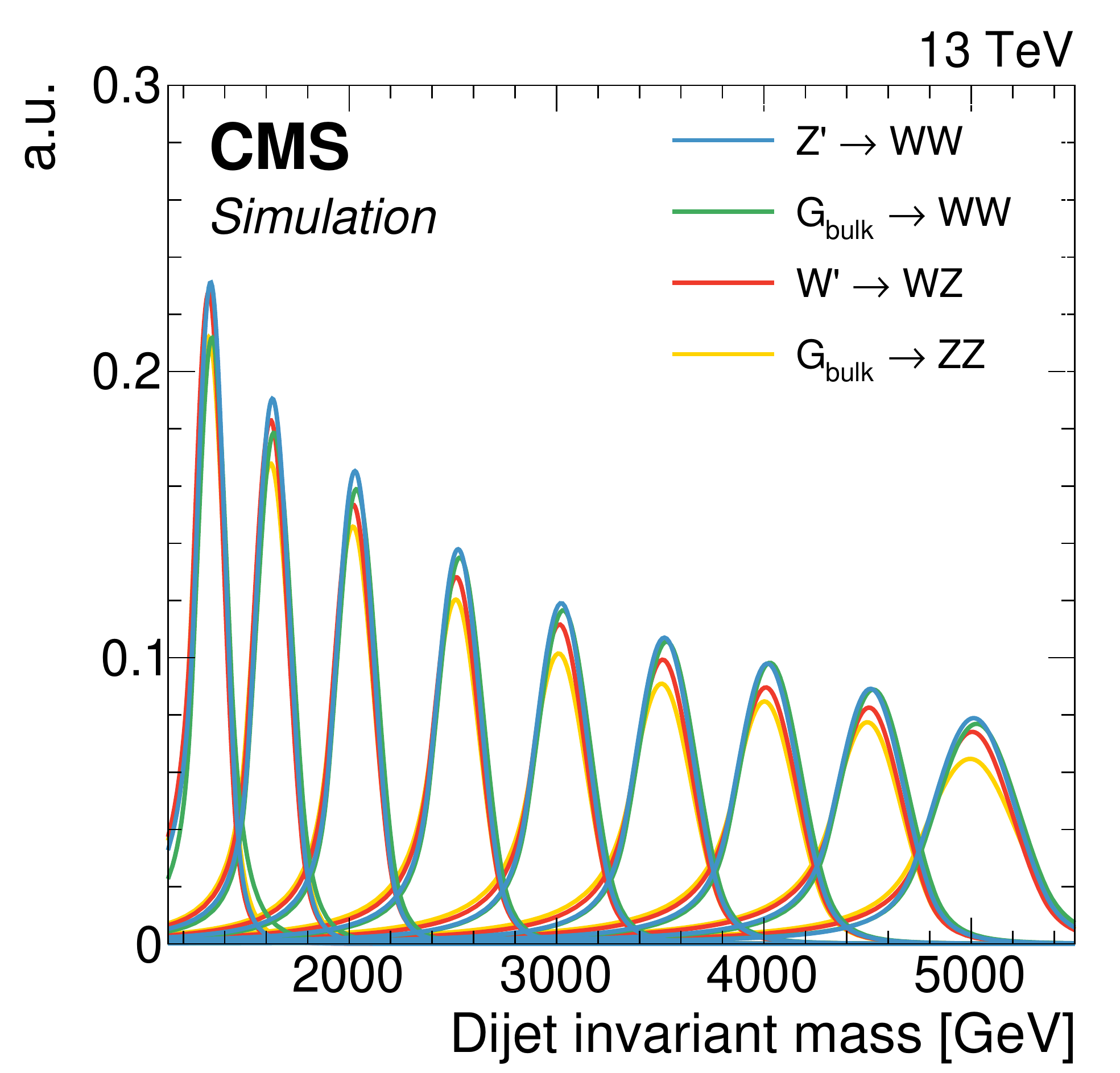}
\includegraphics[width=0.49\textwidth]{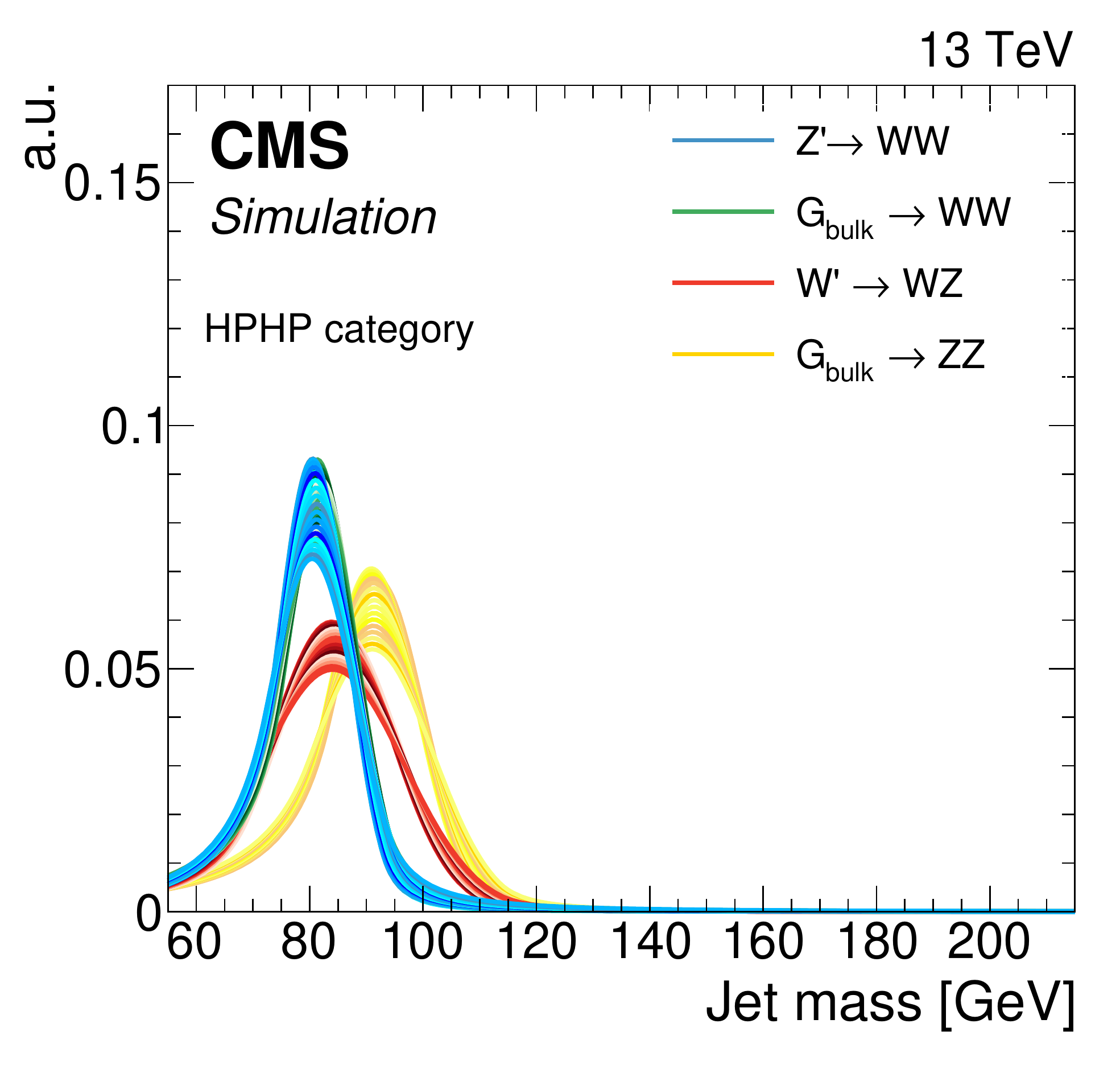}
\caption{The final \MVV (\cmsLeft) and \MJO (\cmsRight) signal shapes extracted from the parameterization of the dCB function. The same \MVV shapes are used for both purity categories. The jet mass distributions are shown for a range of resonance masses between 1.2 and 5.2\TeV for one of the two jets in the events in the HPHP category. Because the jets are labelled randomly, the jet mass distributions for the second jet are essentially the same as the one shown. The distributions for a $\BulkG$ decaying to \PW{}\PW{} have the same shapes as those for the \PZpr signal and are therefore not visible.}
\label{fig:MVVfromjson}
\end{figure}

\begin{figure}[htbp]
\centering
\includegraphics[width=0.49\textwidth]{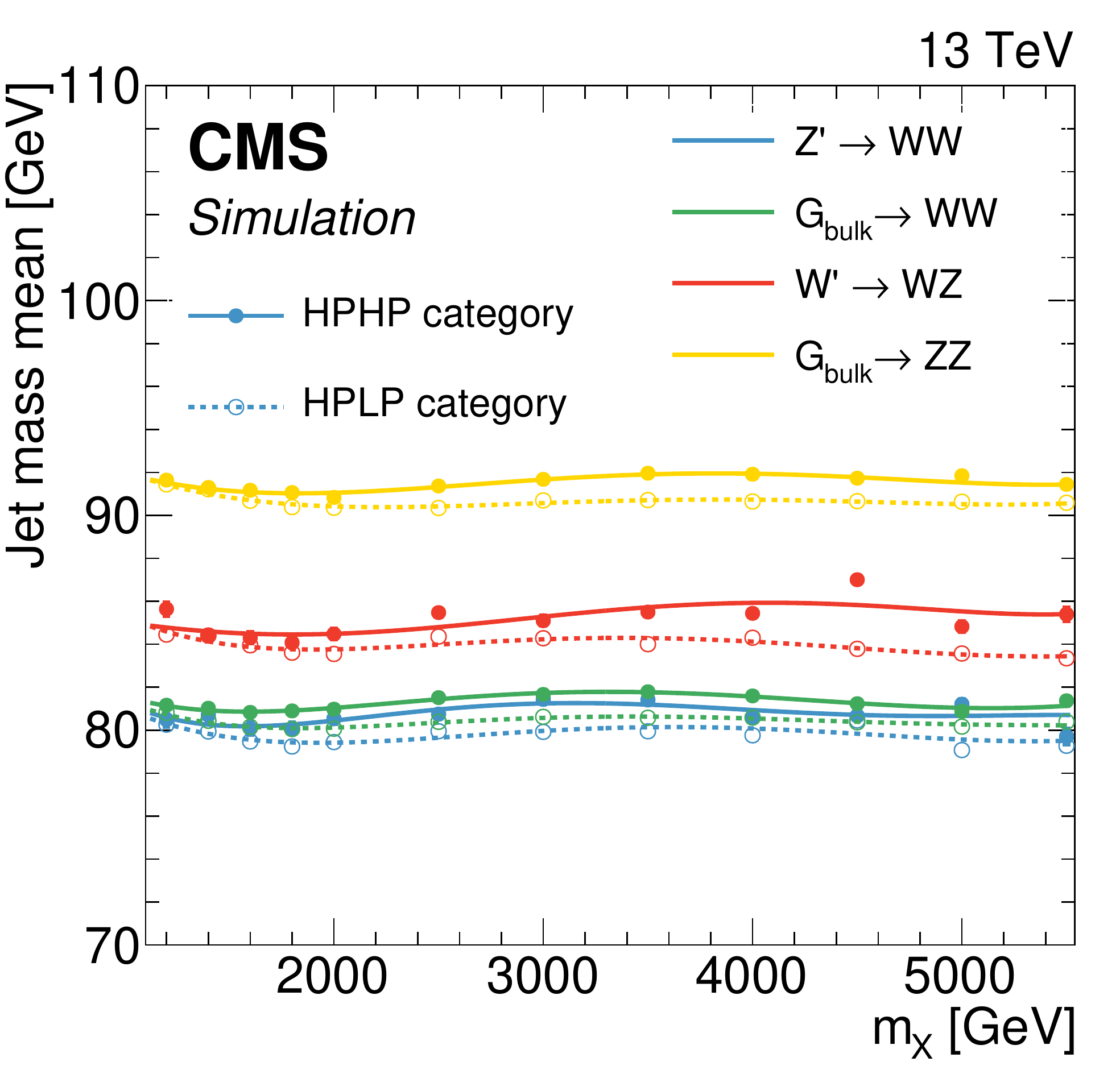}
\includegraphics[width=0.49\textwidth]{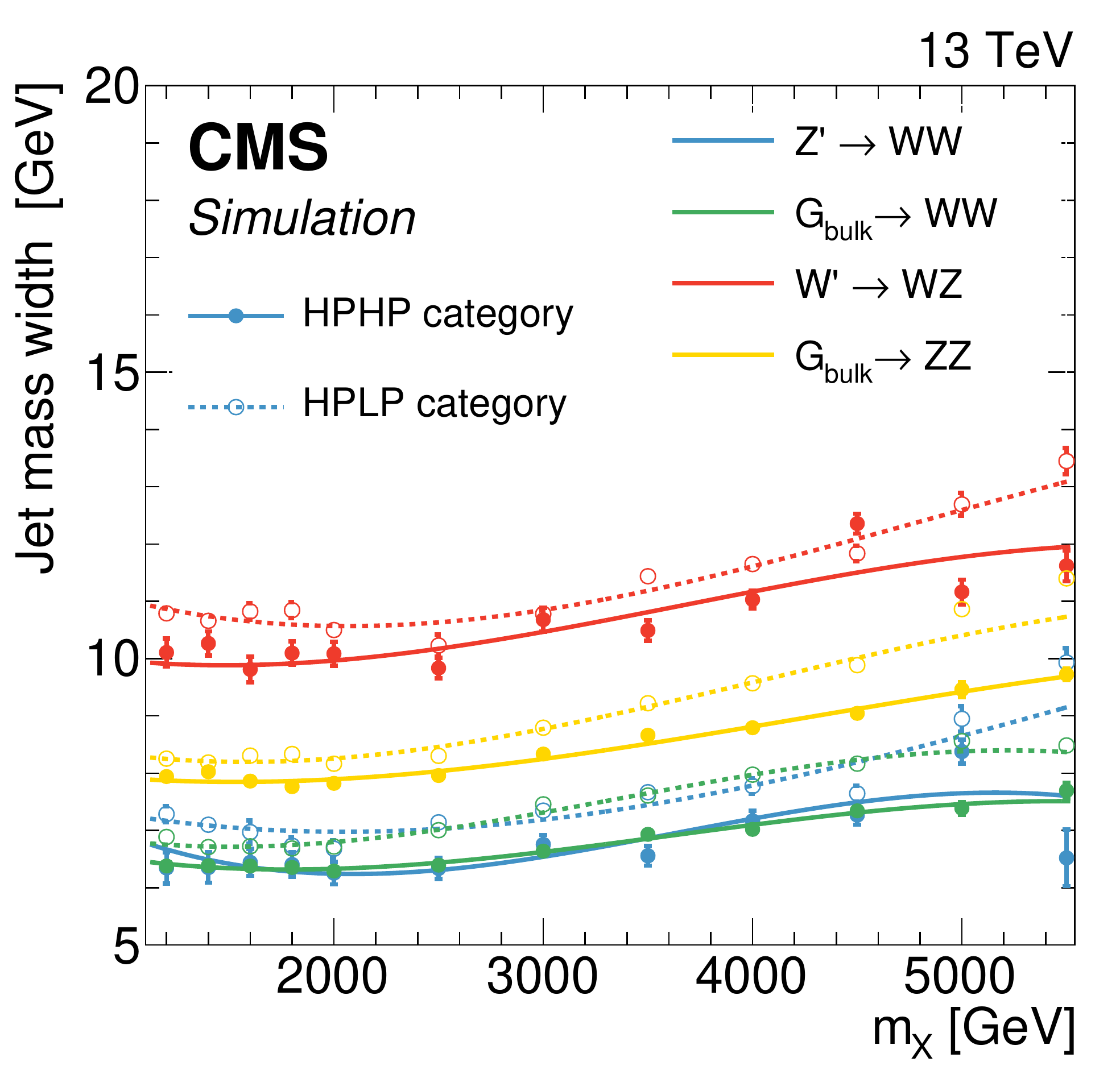}
\caption{The mass scale (\cmsLeft) and resolution (\cmsRight) of the jet as a function of \MX, obtained from the mean and width of the dCB function used to fit the jet mass spectrum. The HPHP (solid lines) and HPLP (dotted lines) categories are shown for different signal models. The distributions are only shown for one of the two jets in the event, since the distributions for the second jet are essentially the same.}
\label{fig:sigscaleres}
\end{figure}

The signal shape in three dimensions is defined as a product of the shape of the resonance mass and the jet masses:
\begin{linenomath}
\ifthenelse{\boolean{cms@external}}
{ 
\begin{multline*}
P_\mathrm{sig}(\MVV,\MJO,\MJT|\overline{\theta}^\mathrm{s}(\MX) ) = P_{\PV{}\PV{}}(\MVV|\overline{\theta}_1^\mathrm{s}(\MX))\\
\times P_{\mathrm{j1}}(\MJO|\overline{\theta}_2^\mathrm{s}(\MX)) \, P_{\mathrm{j2}}(\MJT|\overline{\theta}_3^\mathrm{s}(\MX)).
\end{multline*}
} 
{ 
\begin{equation*}
P_\mathrm{sig}(\MVV,\MJO,\MJT|\overline{\theta}^\mathrm{s}(\MX) ) = P_{\PV{}\PV{}}(\MVV|\overline{\theta}_1^\mathrm{s}(\MX)) \, P_{\mathrm{j1}}(\MJO|\overline{\theta}_2^\mathrm{s}(\MX)) \, P_{\mathrm{j2}}(\MJT|\overline{\theta}_3^\mathrm{s}(\MX)).
\end{equation*}
} 
\end{linenomath}
The shapes for \MVV, \MJO, and \MJT are parameterized independently as a function of the hypothesized mass (\MX{}) of a new particle and a set of general probability density function (pdf) parameters $\overline{\theta}^\mathrm{s}=(\overline{\theta}_1^\mathrm{s}, \overline{\theta}_2^\mathrm{s}, \overline{\theta}_3^\mathrm{s})$ that depend on \MX. The parameters $\overline{\theta}^\mathrm{s}$ denote for instance the mean and width of the analytic function chosen to model the signal distributions. The \MJ and \MVV distributions can be treated as uncorrelated since correlations are found to be negligible for the signal. The signal is parameterized by fitting the simulated resonance mass and jet mass line shapes for each \MX, extracting the quantities, and then interpolating these to intermediate values of the resonance mass. $P_{\mathrm{j1}}$ and $ P_{\mathrm{j2}}$ are fitted and parameterized separately from each other using different sets of $\overline{\theta}^\mathrm{s}$, although they are effectively identical because of the random jet labelling. For the parameterization of the resonance mass \MVV and the \MJ masses, double-sided Crystal Ball (dCB) functions \cite{Oreglia:1980cs} are used for each \MX.
Each parameter of the dCB is interpolated between different resonance masses using polynomials of a degree sufficient to ensure a smooth shape interpolation for all resonance masses.
 The resulting signal shapes for all signal models are shown in Fig.~\ref{fig:MVVfromjson} for the dijet invariant mass (\cmsLeft) and the mass of jet-1 in the HPHP category (\cmsRight). Because of the random jet labelling the distribution for jet-2 is effectively identical to that shown for jet-1. The jet mass scale and resolution as a function of the dijet invariant mass are extracted from the mean and width of the dCB function. The mass scale and resolution are shown in Fig.~\ref{fig:sigscaleres} after the full HPHP (HPLP) analysis selections have been applied. The jet mass resolution increases about 3\% from the lowest to the highest resonance mass, while its scale is found to be stable. The mean of the dijet invariant mass distributions is consistent with the mass of the resonance \MX, as seen in Figure~\ref{fig:MVVfromjson} \cmsLeft.

\subsection{Background modelling}
\label{sec:backgroundestimation}

\subsubsection{Nonresonant background}
\label{subsec:QCDtemplates}

As mentioned above, previous versions of this analysis estimate the QCD multijet background by a parametric fit to the data in the $\MVV$ signal region. The fit is well-constrained by highly populated bins with small statistical uncertainties at low $\MVV$, but is less constrained for high values of $\MVV$.
This method allows the incorporation of additional information in the fit by modelling the correlations between the jet mass and the dijet invariant mass for SM background processes, which were not explicitly studied in the past.
It is important to note that the correlations between $\MJ$ and $\MVV$ have to be modelled for the QCD multijet background, as opposed to the signals negligible correlations due to its localization in the three-dimensional space.
In this analysis, we build a three dimensional background model starting from simulation. Since the size of the simulated samples is limited, we start from particle-level information and use a ``forward-folding" kernel approach that is similar to the technique presented in Ref.~\cite{Cranmer:2000du} and used in Ref.~\cite{Sirunyan:2018iff}. Finally, we incorporate sufficient nuisance parameters into the fit to account for any discrepancies between data and simulation. In order to model the QCD multijet background in the 3D \MVV-\MJO-\MJT hyperplane, we use the following conditional product:
\begin{linenomath}
\ifthenelse{\boolean{cms@external}}
{ 
\begin{multline*}
P_\mathrm{QCD}(\MVV,\MJO,\MJT |\overline{\theta}^{\mathrm{QCD}}) = P_{\PV{}\PV{}}(\MVV|\overline{\theta}_1^{\mathrm{QCD}})\\
\times P_\mathrm{cond,1}(\MJO|\MVV,\overline{\theta}_2^{\mathrm{QCD}}) \, P_\mathrm{cond,2}(\MJT|\MVV,\overline{\theta}_3^{\mathrm{QCD}}).
\end{multline*}
} 
{ 
\begin{equation*}
P_\mathrm{QCD}(\MVV,\MJO,\MJT |\overline{\theta}) = P_{\PV{}\PV{}}(\MVV|\overline{\theta}_1^{\mathrm{QCD}}) \, P_\mathrm{cond,1}(\MJO|\MVV,\overline{\theta}_2^{\mathrm{QCD}}) \, P_\mathrm{cond,2}(\MJT|\MVV,\overline{\theta}_3^{\mathrm{QCD}}).
\end{equation*}
} 
\end{linenomath}

Since the jet mass is correlated with the jet \PT for the QCD multijet background, its shape is required to be modelled conditionally as a function of $\MVV$ for both jets. Two two-dimensional (2D) templates (denoted as $P_\mathrm{cond,1}$ and $P_\mathrm{cond,2}$) are modelled for the two jet masses separately, containing different jet mass shapes in bins of $\MVV$.
The $\MVV$ distribution is computed as a 1D pdf. The parameter sets denoted by $\overline{\theta}^{\mathrm{QCD}}$ represent the nuisance parameters in each pdf.

To build the 2D conditional templates, $P_\mathrm{cond,1}$ and $P_\mathrm{cond,2}$, each available particle-level event is smoothed with a 2D Gaussian kernel, where each 2D kernel links the particle-level event quantities to the reconstruction level. Thus each simulated event contributes a smoothed Gaussian shape to the total conditional pdf.
The Gaussian kernel depends on the dijet invariant mass scale and resolution, as well as the jet mass scale and resolution.
The $\MJ$ and $\MVV$ scale and resolution are extracted from a Gaussian fit to either $\MJ(\mathrm{reco})/\MJ(\mathrm{gen})$ or $\MVV(\mathrm{reco})/\MVV(\mathrm{gen})$, in bins of particle-level jet \PT.

The mass scale and resolution are then used to populate the conditional 2D histogram. Each generated event $i$ is smeared with a 2D Gaussian kernel,
\begin{linenomath}
\ifthenelse{\boolean{cms@external}}
{ 
\begin{multline*}
k(\MJ,\MVV) =\frac{w_i}{2\pi r_{\MVV,i} r_{\MJ,i}}\\
\times \exp \left [-\frac{1}{2} \left ( \frac{\MVV-s_{\MVV,i}}{r_{\MVV,i}} \right)^2-\frac{1}{2} \left ( \frac{\MJ-s_{\MJ,i}}{r_{\MJ,i}} \right)^2 \right ],
\end{multline*}
} 
{ 
\begin{equation*}
k(\MJ,\MVV) =\frac{w_i}{2\pi r_{\MVV,i} r_{\MJ,i}} \exp \left [-\frac{1}{2} \left ( \frac{\MVV-s_{\MVV,i}}{r_{\MVV,i}} \right)^2 -\frac{1}{2} \left ( \frac{\MJ-s_{\MJ,i}}{r_{\MJ,i}} \right)^2 \right ],
\end{equation*}
} 
\end{linenomath}
where $s_{i}$ and $r_{i}$ are the scale and the resolution derived in the previous step, and $w_i$ is a product of event weights accounting for the normalization effects such as the individual sample production cross sections. In this way, the jet mass in generated events is scaled and smeared according to the evaluated scale and resolution, and a 2D histogram is filled with smooth Gaussian shapes. According to this procedure the jet mass (\mjj) resolution is about 7--10\% (3--7\%) of the mass of the generated jet, depending on its \PT.
This procedure is performed separately for \MJO and \MJT however the two resulting templates $P_\mathrm{cond,1}$ and $P_\mathrm{cond,2}$ are essentially the same because of the random jet labels. Finally, we interpolate the 2D histogram in order to have valid values of the pdf in all $\MVV$ bins. Starting from the histogram, coarsely binned in $\MVV$, for each $\MJ$ bin a spline is fitted over all $\MVV$ bins. The spline is then used to interpolate values of the histogram for all final $\MVV$ bins, resulting in a 2D histogram with the desired binning.

To build the 1D template for the dijet invariant mass, $P_{\PV{}\PV{}}$, a 1D Gaussian kernel is constructed starting from particle-level quantities where, for each MC event, a Gaussian probability distribution, rather than a single point representing the mean, contributes to the total 1D pdf using the same procedure as for the 2D templates.

\begin{figure*}[htbp]
\centering
\includegraphics[width=0.45\textwidth]{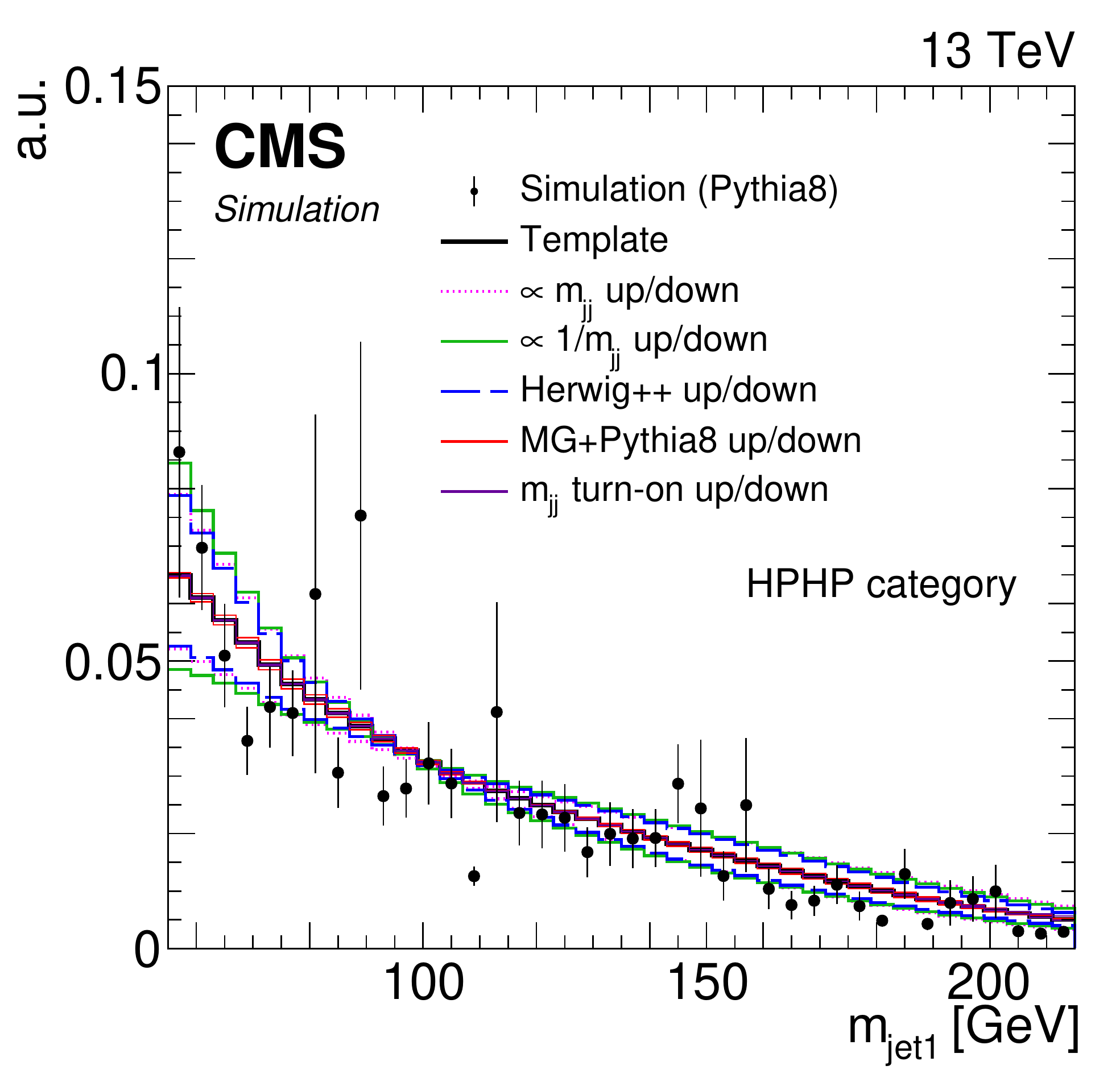}
\includegraphics[width=0.45\textwidth]{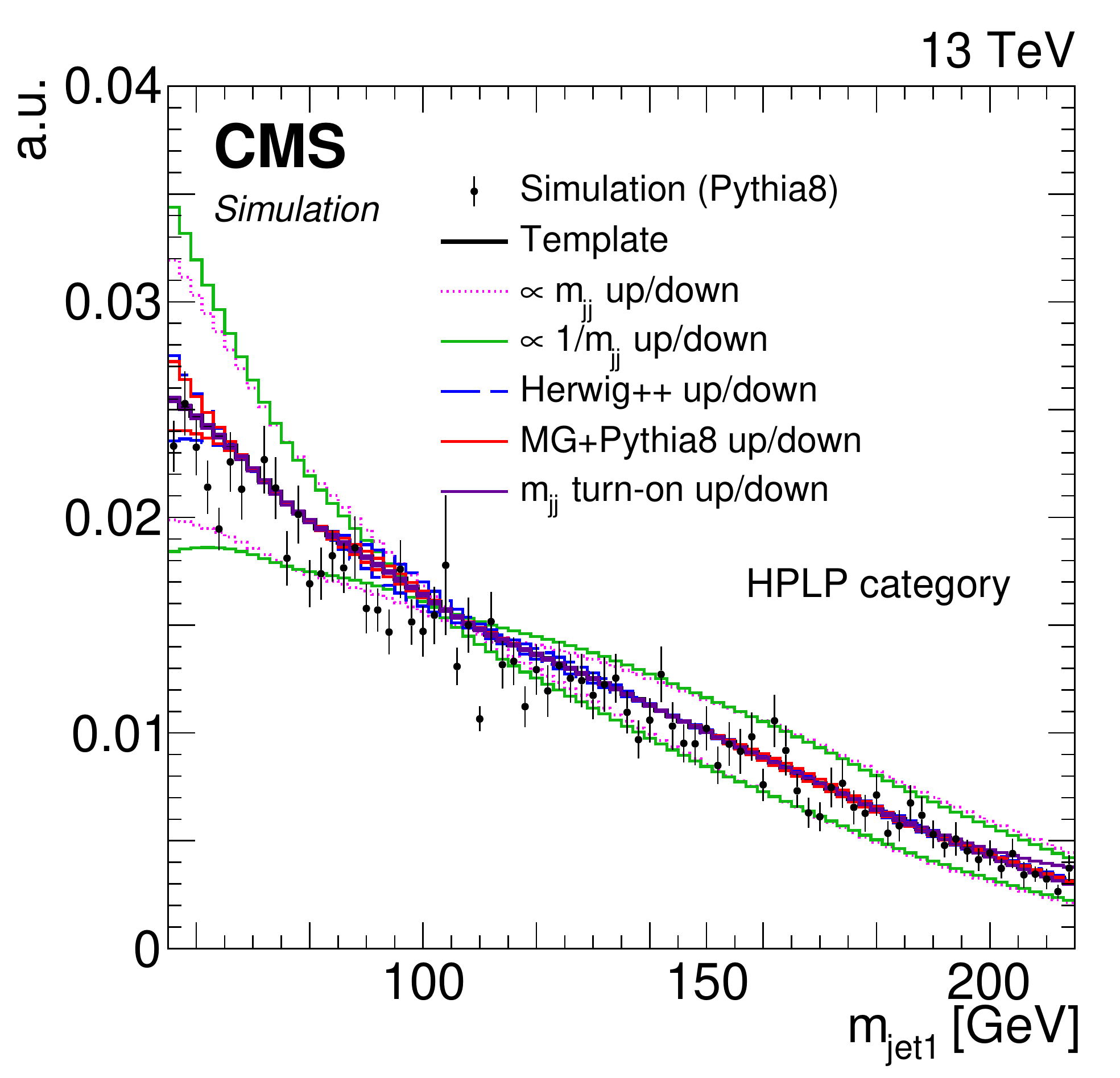}\\
\includegraphics[width=0.45\textwidth]{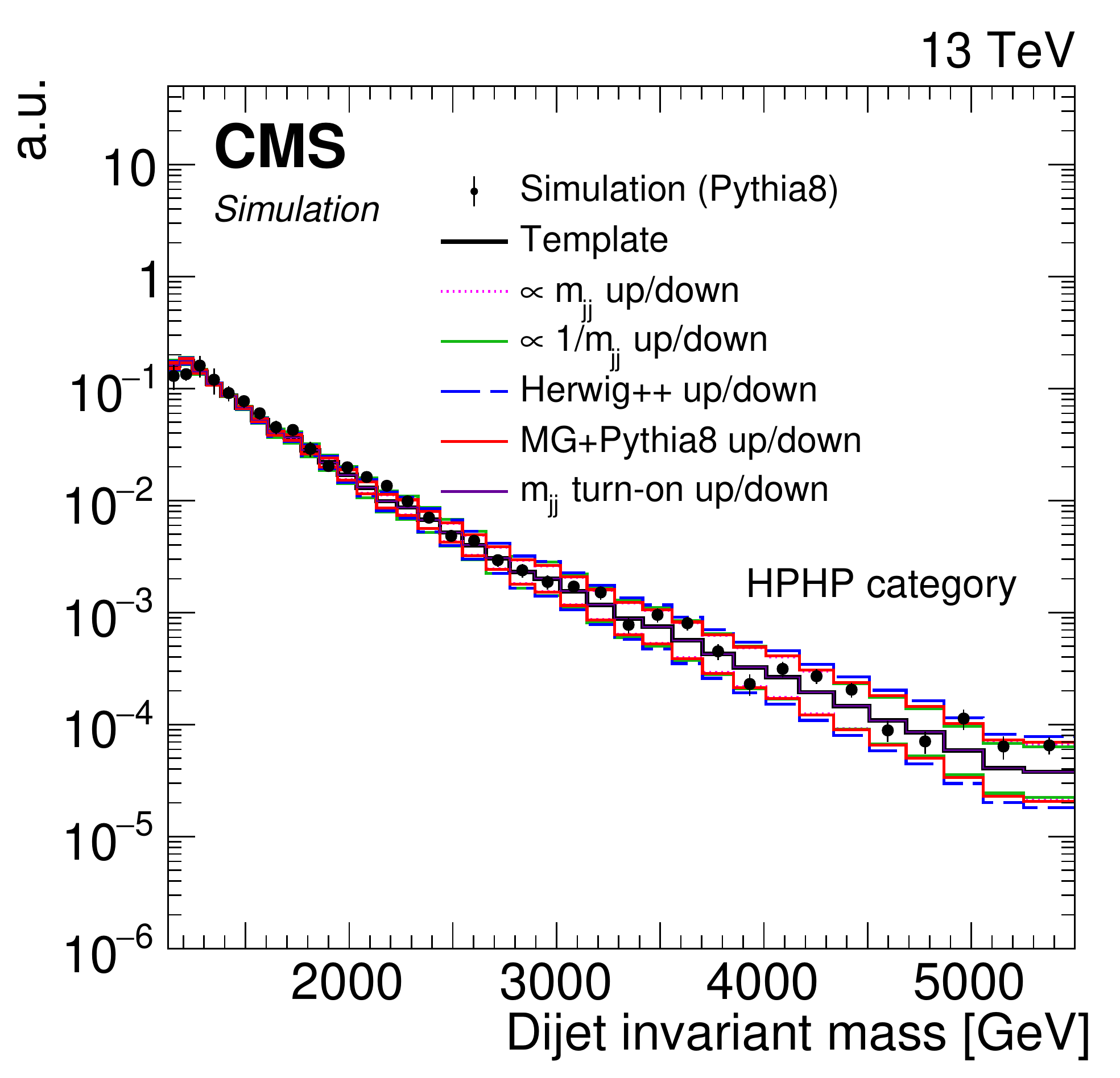}
\includegraphics[width=0.45\textwidth]{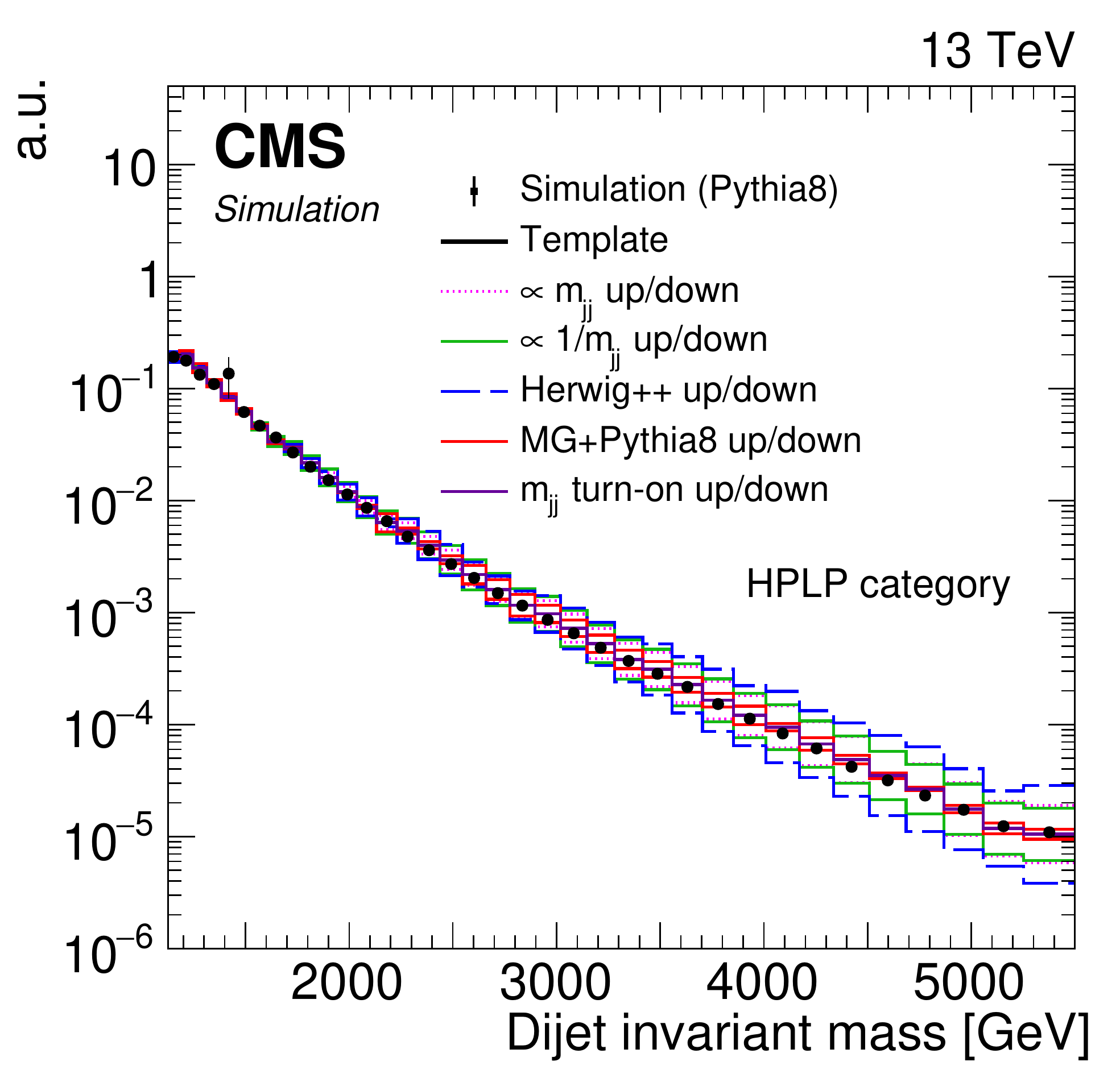}
\caption{Nominal QCD multijet simulation using \PYTHIA{8} (data points) and three-dimensional pdfs derived using a forward-folding kernel approach (black solid line), shown together with the five alternate shapes that are added to the multi-dimensional fit as shape nuisance parameters. The shapes for the high-purity (left) and low-purity (right) categories obtained with the 2017 simulation are shown for the projection on \MJO (upper) and \MVV (lower). The projection on \MJT is omitted since it is equivalent to the \MJO projection except for statistical fluctuations. The pdfs and the simulations shown are normalized to unity. The normalization uncertainty of 50\% is not shown. The distributions for 2016 simulations are similar.}
\label{fig:sys_HPHP}
\end{figure*}

Because of the low number of events in the HPHP category, the 3D kernel derived in the HPLP category, which has a similar shape, is used to model the HPHP background. This is done by adapting the templates derived in the HPLP category to the HPHP category in the QCD multijet simulation through a fit, and then by using the corresponding post-fit distribution as the nominal template for the HPHP category. The free parameters in the fit are the alternate shapes proportional to \MVV, \MJ,  1/\MVV, and 1/\MJ, as listed in Section~\ref{sec:systematicuncertainties}.
The projections on the three different axes of the final 3D pdf, in bins of $\MVV$ or $\MJ$, are shown in Fig.~\ref{fig:sys_HPHP}, compared to the spectra obtained using bare QCD multijet simulation events. Good agreement is observed, and any residual discrepancies are covered by the systematic shape uncertainties described in Section~\ref{sec:systematicuncertainties} and also shown in Fig.~\ref{fig:sys_HPHP}. Repeating the template building process and performing fits to a control region in data where both jets fail the high-purity requirement confirms the validity of this approach. In addition, the adaptability of the method was further confirmed by fitting a QCD multijet background generated at NLO with \POWHEG{}.

\subsubsection{Resonant background}

The resonant background is defined as all SM processes containing at least one jet originating from a genuine \PW\ or \PZ\ boson decay. It is dominated by \PV{}+jets events, with a minor contribution from \ttbar{} production and an inconsequential contribution from SM $\PV{}\PV{}$ production, that is absorbed into the \PV{}+jets modelling. As the labelling of each jet is arbitrary, each jet mass distribution contains two contributions: a resonant part consisting of genuine vector-boson jets, peaking around the \PW\ or \PZ\ boson mass; and a nonresonant part, composed of mistagged jets originating from a prompt quark or a gluon. These two contributions are modelled separately for each jet mass dimension.
A 3D pdf for the resonant backgrounds, $P_\mathrm{\PV{}+jets}$, is built as a product of three 1D pdfs as follows:
\begin{linenomath}
\ifthenelse{\boolean{cms@external}}
{ 
\begin{multline*}
P_\mathrm{\PV{}+jets} \left(\MJO, \MJT, \MVV |\overline{\theta} \right) = 0.5 \, (P_{\PV{}\PV{}} ( \MVV |\overline{\theta}_1  )\\
\times P_\text{res} ( \MJO |\overline{\theta}_2 ) \, P_\text{nonres} (\MJT |\overline{\theta}_3 )) + 0.5 \, ( P_{\PV{}\PV{}} ( \MVV |\overline{\theta}_1 )\\
\times P_\text{res} ( \MJT |\overline{\theta}_2 ) \, P_\text{nonres} (\MJO |\overline{\theta}_3)).
\end{multline*}
} 
{ 
\begin{align*}
\begin{split}
P_\mathrm{\PV{}+jets} \left(\MJO, \MJT, \MVV |\overline{\theta} \right) &= 0.5 \, \left( P_{\PV{}\PV{}} ( \MVV |\overline{\theta}_1  ) \, P_\text{res} ( \MJO |\overline{\theta}_2 ) \, P_\text{nonres} (\MJT |\overline{\theta}_3 ) \right) \\
&+ 0.5 \, \left( P_{\PV{}\PV{}} ( \MVV |\overline{\theta}_1 ) \, P_\text{res} ( \MJT |\overline{\theta}_2 ) \, P_\text{nonres} (\MJO |\overline{\theta}_3) \right).
\end{split}
\end{align*}
} 
\end{linenomath}
The resonant mass shape $P_\text{res}$ is derived by fitting a dCB function to the simulated jet mass spectrum, performed separately for \MJO and \MJT. The resonant events are separated from the nonresonant ones when building the pdfs by requiring that there is a generated boson in a cone of $\Delta R=0.8$ around the reconstructed merged jet.
The nonresonant component of the jet mass shape is fitted separately with a Gaussian function. The contributions of \PW{}+jets and \ttbar{} production are considered as one combined background shape, because both have a resonant peak around the \PW{}-boson mass, while the \PZ{}+jets background contribution is considered separately.
The top mass peak does not need to be modelled since the overall contribution of \ttbar{} events is less than 2\%.
The nonresonant dijet invariant mass shape of the \PV{}+jets backgrounds, $P_{\PV{}\PV{}}$, is modelled with a one dimensional kernel, in the same way as the dijet invariant mass shape of the QCD multijet background.

\section{Systematic uncertainties}
\label{sec:systematicuncertainties}

\subsection{Systematic uncertainties in the background estimation}

Uncertainties in the QCD multijet background shape are included in the fit using alternative pdfs derived with the template-building method described in Section~\ref{subsec:QCDtemplates}.
We define five nuisance parameters that vary the shape, each of the parameters corresponding to an upward and a downward variation of alternative shapes that simultaneously affect all three dimensions.
The first nuisance parameter accounts for a variation of the underlying \PT spectrum, and the two corresponding mirrored templates are obtained by applying up and down variations
of the expected yields to each bin along the two jet masses and \MVV by a quantity proportional to \MJ and \MVV.
The second nuisance parameter is a variation of the mass scale, and is taken into account through two mirrored alternative shapes obtained by applying up and down variations
of each bin content along the two jet masses and \MVV by a quantity proportional to 1/\MJ and 1/\MVV.
Two additional alternative shapes that simultaneously affect resonance mass and jet groomed mass are also added in order to take into account differences in MC generation and modelling of the parton shower. These alternative templates are derived using the \HERWIG{++} and \MADGRAPH{}+\PYTHIA{8} QCD multijet simulation. This allows us to include all known background variations into the fit.
For events with a large \MJ ($>$175\GeV) and low \MVV($<$1200\GeV), there is an expected turn-on due to the trigger thresholds. Therefore, an additional shape uncertainty parameterizing any discrepancy between the 3D template and the QCD multijet simulation is added to the fit. Note that this shape uncertainty only affects this particular region, which is far from where a diboson signal, as relevant for this analysis, is expected.
The nuisance parameters associated with these alternative shapes are constrained using Gaussian pdfs in the fit, with the pre-fit values chosen in order to cover any differences between data and simulation observed in the control regions.
The alternative shapes described above are shown in Fig.~\ref{fig:sys_HPHP}.

A similar procedure is used for the \PV{}+jets background, adding two alternative shapes to the $\MVV$ templates derived by a variation proportional to $\MVV$ and $1/\MVV$. The resonant jet mass shapes for this background are subject to the same uncertainties as the signal.
The normalizations of the \PV{}+jets and QCD background are obtained directly from simulation and are allowed to vary within 20 and 50\%, respectively.
The same nuisance parameters are used for the fit to 2016 and 2017 data, which reduces the fit complexity while not impacting the result of the fit.

\subsection{Systematic uncertainties in the signal prediction}

{\tolerance=800
The dominant uncertainty in the signal selection efficiency arises from uncertainties in the boson tagging efficiency. The effect of this uncertainty is evaluated per jet and assumed to be fully correlated between both jets in the event.
The \PW\ boson tagging efficiency scale factor is fully anticorrelated between the HPHP and HPLP categories (3--10\%), and fully correlated between signal and \PV{}+jets backgrounds. The \PT-dependence uncertainty in the scale factor arises from the extrapolation to higher \PT{}'s of the \PW\ boson tagging efficiency scale factors, which are measured in \ttbar{} events where the jet has a \pt around 200\GeV. This uncertainty is estimated in signal simulation, and is based on the difference in tagging efficiency between the samples matched and showered either with \PYTHIA{} or with \HERWIG{++}, as a function of \pt, relative to the difference at 200\GeV. This is considered as correlated between the $\nsubjDDT$ categories, and is given as $6\,(7) \% \ln(\PT/300\,(\GeVns{}))$ for the HPHP (HPLP) categories.
The shape uncertainties in the jet masses are considered fully correlated between signal and \PV{}+jets backgrounds, allowing the data to constrain these parameters. These affect the mean and the width of the signal and \PV{}+jets pdfs. All uncertainties entering the fit are listed in Table~\ref{tab:systematics}.
\par}

\begin{table*}[htbp]
  \centering
  \topcaption{Summary of the systematic uncertainties and their impact the affected quantities. Numbers in parentheses correspond to uncertainties for the 2016 analysis if these differ from those for 2017. Dashes indicate shape variations that cannot be described by a single parameter, and are discussed in the text.}
  \cmsTable{
  \begin{tabular}{llcc}
    \hline
    Source                        & Relevant quantity      & HPHP unc. (\%)  & HPLP unc. (\%)\\
    \hline
    PDFs                      & Signal yield                  & \multicolumn{2}{c}{3}\\
    \PW boson tagging efficiency      & Signal + \PV{}+jets yield       & 25 (21)              & 13 (11) \\
    \PW boson tagging \PT{} dependence  & Signal + \PV{}+jets yield       & 8--23              & 9--25 \\
    Integrated luminosity     & Signal + \PV{}+jets yield       & \multicolumn{2}{c}{2.3 (2.6)}\\
    QCD normalization         & Background yield              & \multicolumn{2}{c}{50}\\
    \PW{}+jets normalization      & Background yield              & \multicolumn{2}{c}{20}\\
    \PZ{}+jets normalization       & Migration                     & \multicolumn{2}{c}{20}\\
    PDFs                      & Signal \MVV/\MJ mean and width         & \multicolumn{2}{c}{$<$1}\\
    Jet energy scale          & Signal \MVV mean            & \multicolumn{2}{c}{2}\\
    Jet energy resolution     & Signal \MVV width             & \multicolumn{2}{c}{5}\\
    Jet mass scale            & Signal + \PV{}+jets \MJ mean   & \multicolumn{2}{c}{2}\\
    Jet mass resolution       & Signal + \PV{}+jets \MJ width   & \multicolumn{2}{c}{8}\\
    QCD \HERWIG{++}           & QCD shape                     & \multicolumn{2}{c}{\NA}\\
    QCD \MADGRAPH{}+\PYTHIA{8}& QCD shape                     &  \multicolumn{2}{c}{\NA}\\
    \PT{} variations            & QCD shape                     &  \multicolumn{2}{c}{\NA}\\
    Scale variations          & QCD shape                     &  \multicolumn{2}{c}{\NA}\\
    High-\MJ turn-on          & QCD shape                     &  \multicolumn{2}{c}{\NA}\\
    \PT{} variations            & \PV{}+jets \MVV shape             &  \multicolumn{2}{c}{\NA}  \rule[-1.2ex]{0pt}{0pt}\\
    \hline
  \end{tabular}
  \label{tab:systematics}
  }
\end{table*}

\section{Statistical interpretation}
\label{sec:narrow-results}

To test for the presence of narrow resonances decaying to two vector bosons we follow the \CLs prescription, evaluated using asymptotic expressions described in Refs.~\cite{CLs1,Junk:1999kv,Cowan:2010js}.
The limits are computed using a shape analysis of the three dimensional \MVV-\MJO-\MJT spectrum, where the 3D signal and background pdfs obtained above are fitted simultaneously to the data for each signal mass hypothesis and category. The signal and background yields are determined simultaneously in this fit.
Systematic uncertainties are treated as nuisance parameters and profiled in the statistical interpretation using log-normal constraints, while Gaussian constraints are used for shape uncertainties.

\begin{figure*}[htbp]
\centering
\includegraphics[width=0.49\textwidth]{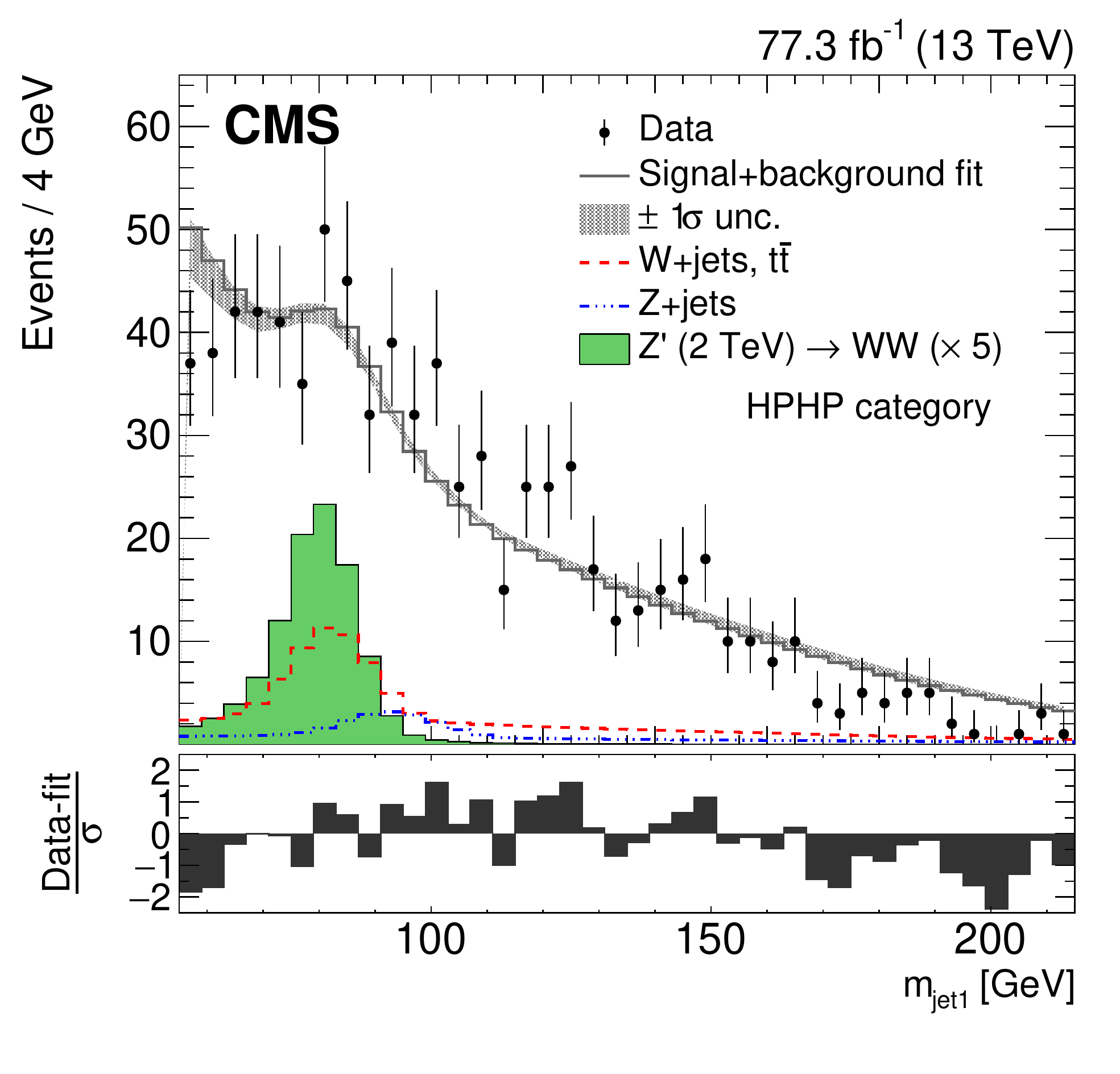}
\includegraphics[width=0.49\textwidth]{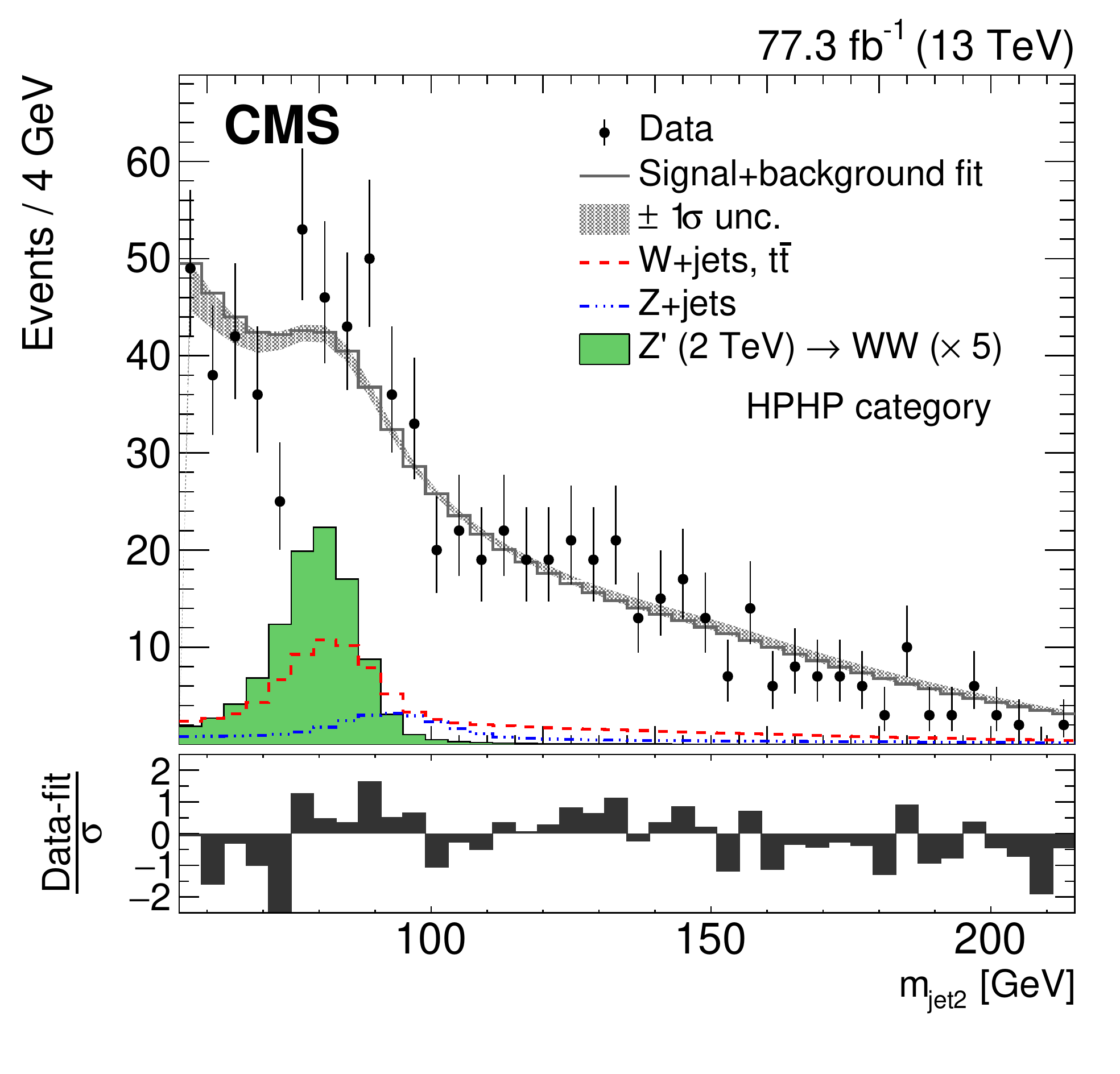}\\
\includegraphics[width=0.49\textwidth]{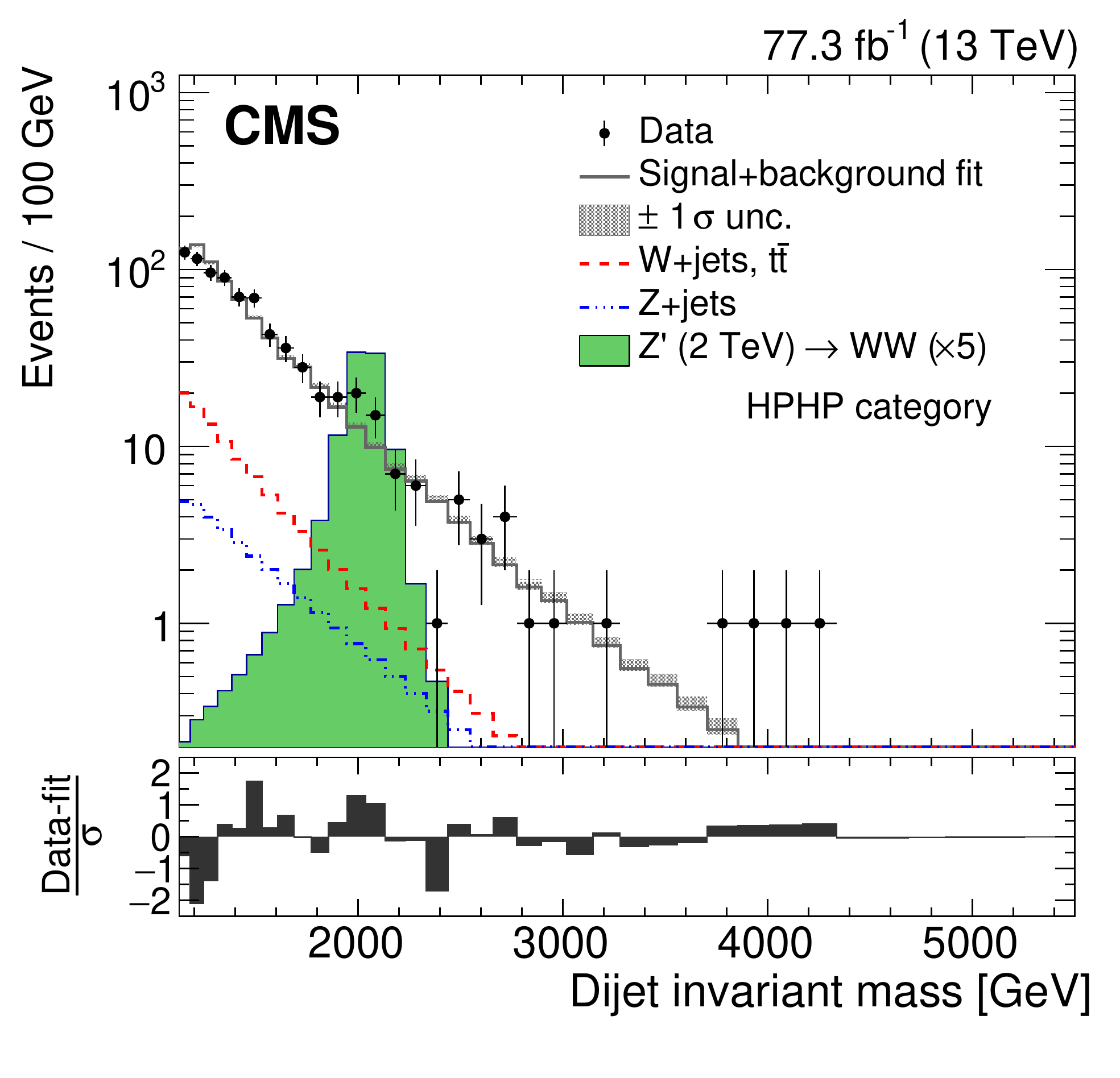}
\caption{For the HPHP category: comparison between the signal+background fit and the data distributions of \MJO (upper left), \MJT (upper right), and \MVV (lower).
The background shape uncertainty is shown as a gray shaded band,  and the statistical uncertainties of the data are shown as vertical bars.
An example of a signal distribution is overlaid, where the number of expected events is scaled by a factor of 5. Shown below each mass plot is the corresponding pull distribution (Data-fit)/$\sigma$, where $\sigma=\sqrt{\sigma_\mathrm{data}^2-\sigma_\mathrm{fit}^2}$ for each bin to ensure a Gaussian pull-distribution, as defined in Ref.~\cite{CDF:AN5776}.}
\label{fig:postfit1}
\end{figure*}

\begin{figure*}[htbp]
\centering
\includegraphics[width=0.49\textwidth]{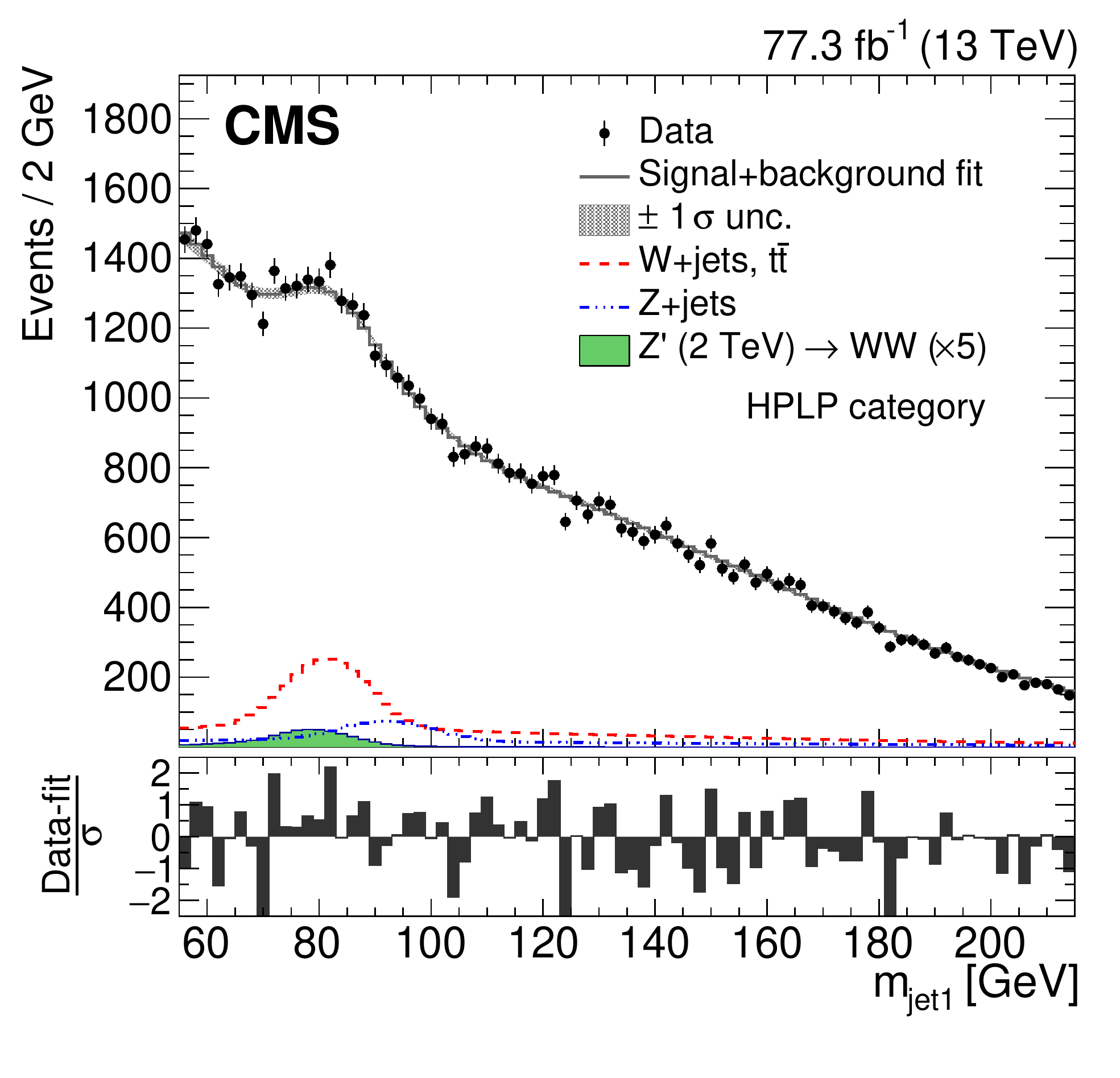}
\includegraphics[width=0.49\textwidth]{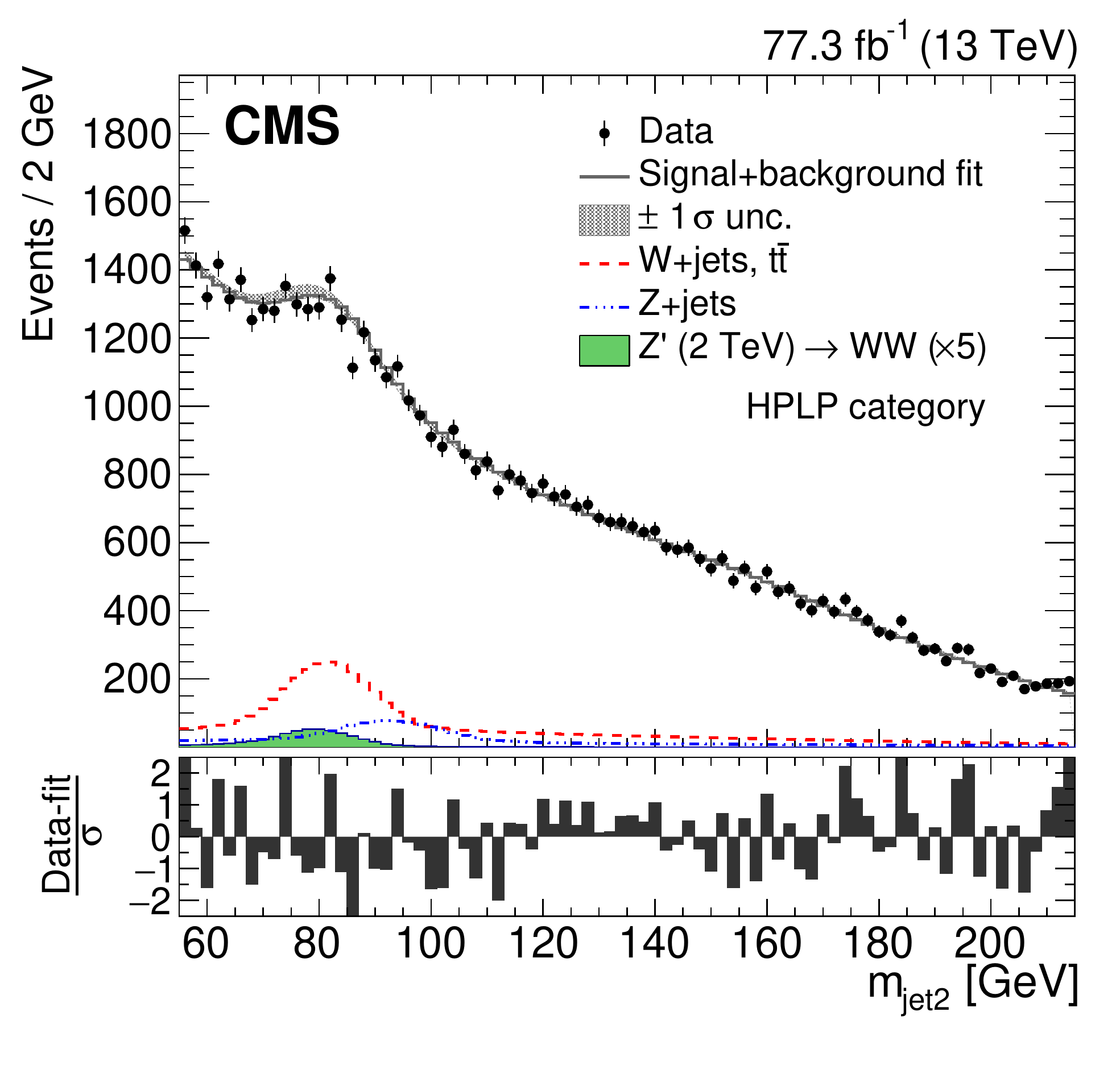}\\
\includegraphics[width=0.49\textwidth]{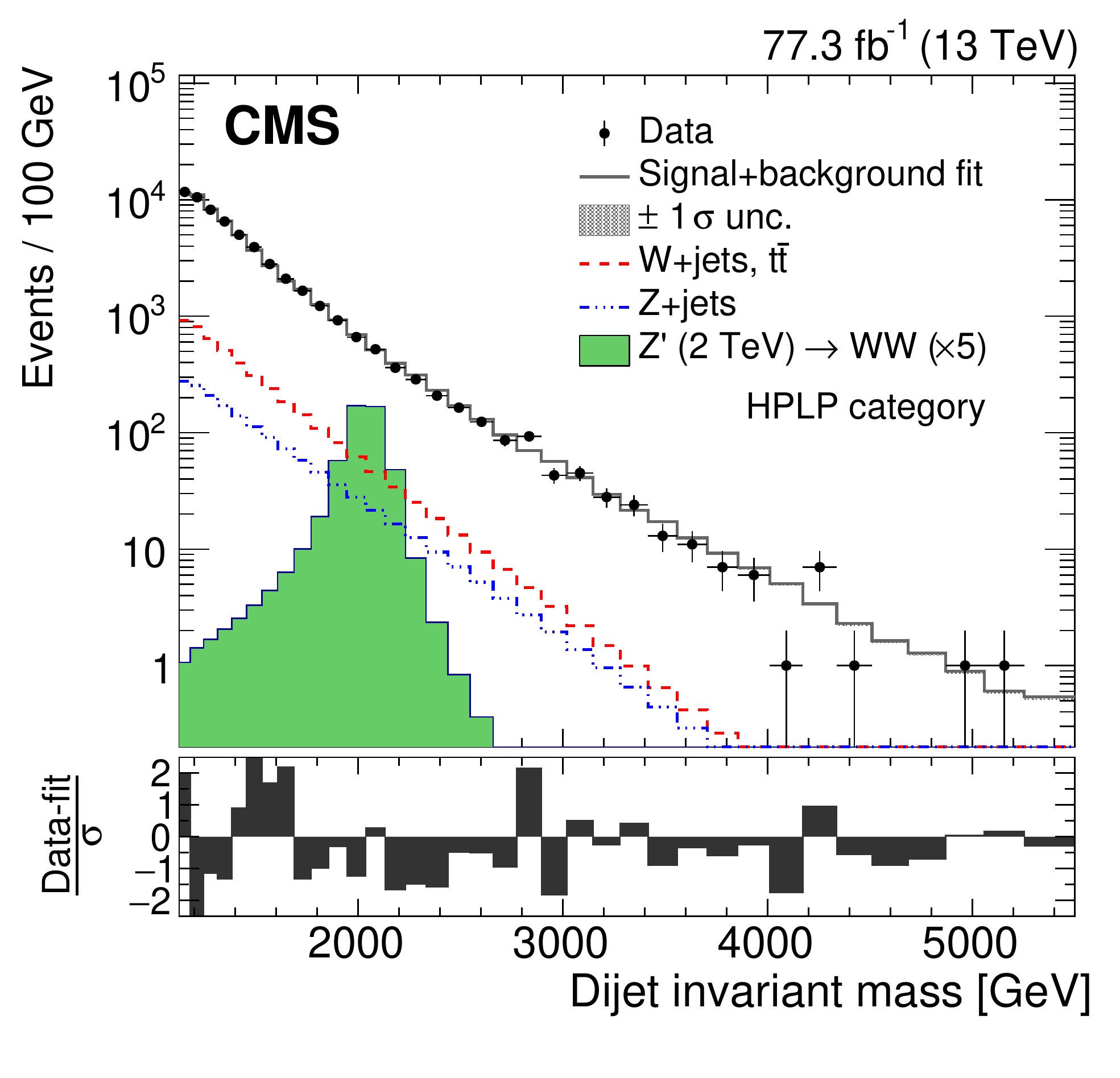}
\caption{For the HPLP category: comparison between the signal+background fit and the data distributions of \MJO (upper left), \MJT (upper right), and \MVV (lower).
The background shape uncertainty is shown as a gray shaded band,  and the statistical uncertainties of the data are shown as vertical bars.
An example of a signal distribution is overlaid, where the number of expected events is scaled by a factor of 5. Shown below each mass plot is the corresponding pull distribution (Data-fit)/$\sigma$, where $\sigma=\sqrt{\sigma_\text{data}^2-\sigma_\text{fit}^2}$ for each bin to ensure a Gaussian pull-distribution, as defined in Ref.~\cite{CDF:AN5776}.}
\label{fig:postfit2}
\end{figure*}

Figures~\ref{fig:postfit1} and \ref{fig:postfit2} show the \MJ and \MVV spectra in data for the high- and low-purity categories, respectively.
The solid gray curves represent the results of the maximum likelihood fit to the data, allowing the signal yields to assume their best fit value, while the lower panels show the corresponding pull distributions, quantifying the agreement between the hypothesis of signal plus background and the data. The resonant background components are shown separately. A signal is superimposed onto all three projections corresponding to a signal yield as expected from the theoretical prediction and the analysis selection efficiency, and scaled by a factor of 5.
The background yields in the signal region extracted from a background-only fit, together with their post-fit uncertainties, are summarised in Table~\ref{tab:ObsEvents} and compared with observations, separately for the two categories. The extracted \PV{}+jets cross sections are found to be compatible with the SM expectations within one standard deviation of the post-fit uncertainties.

\begin{table}[b]
\topcaption{Observed yield and background yields extracted from the background-only fit together with post-fit uncertainties, in the two purity categories.}
\centering
\begin{tabular}{lXX}
\hline
 Category & \multicolumn{1}{c}{HPHP} & \multicolumn{1}{c}{HPLP}\\
 \hline
\PW{}+jets & 100+11 & 4600+200 \\
\PZ{}+jets  & 33+4   & 1580+160 \\
QCD multijets & 650+4     & 51100+300 \\
Predicted total background & 783+12 & 57200+400\\[\cmsTabSkip]
Observed yield & 780+30 & 57230+240\\
\hline
\end{tabular}
\label{tab:ObsEvents}
\end{table}

\begin{figure*}[htbp]
\centering
\includegraphics[width=0.49\textwidth]{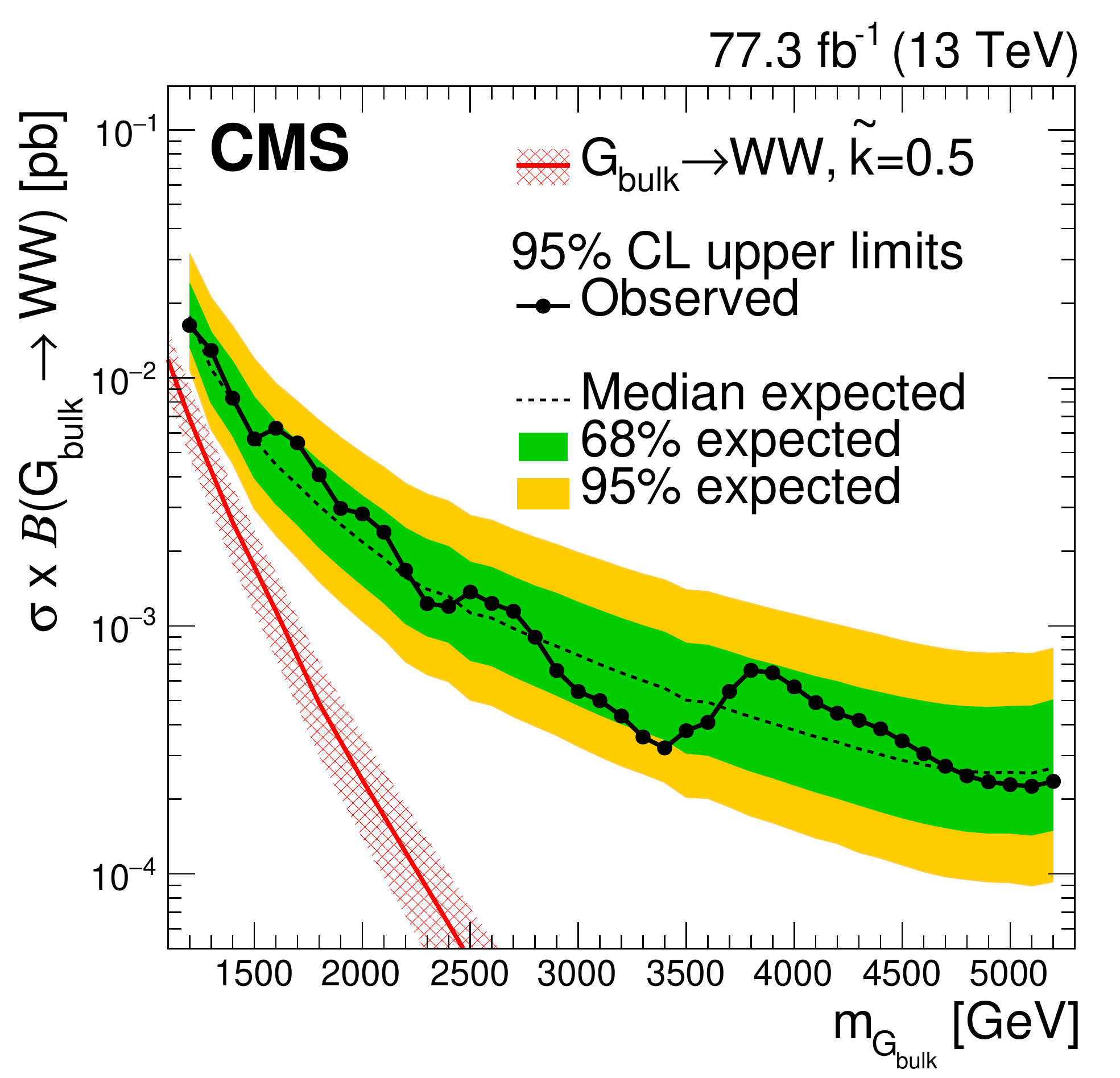}
\includegraphics[width=0.49\textwidth]{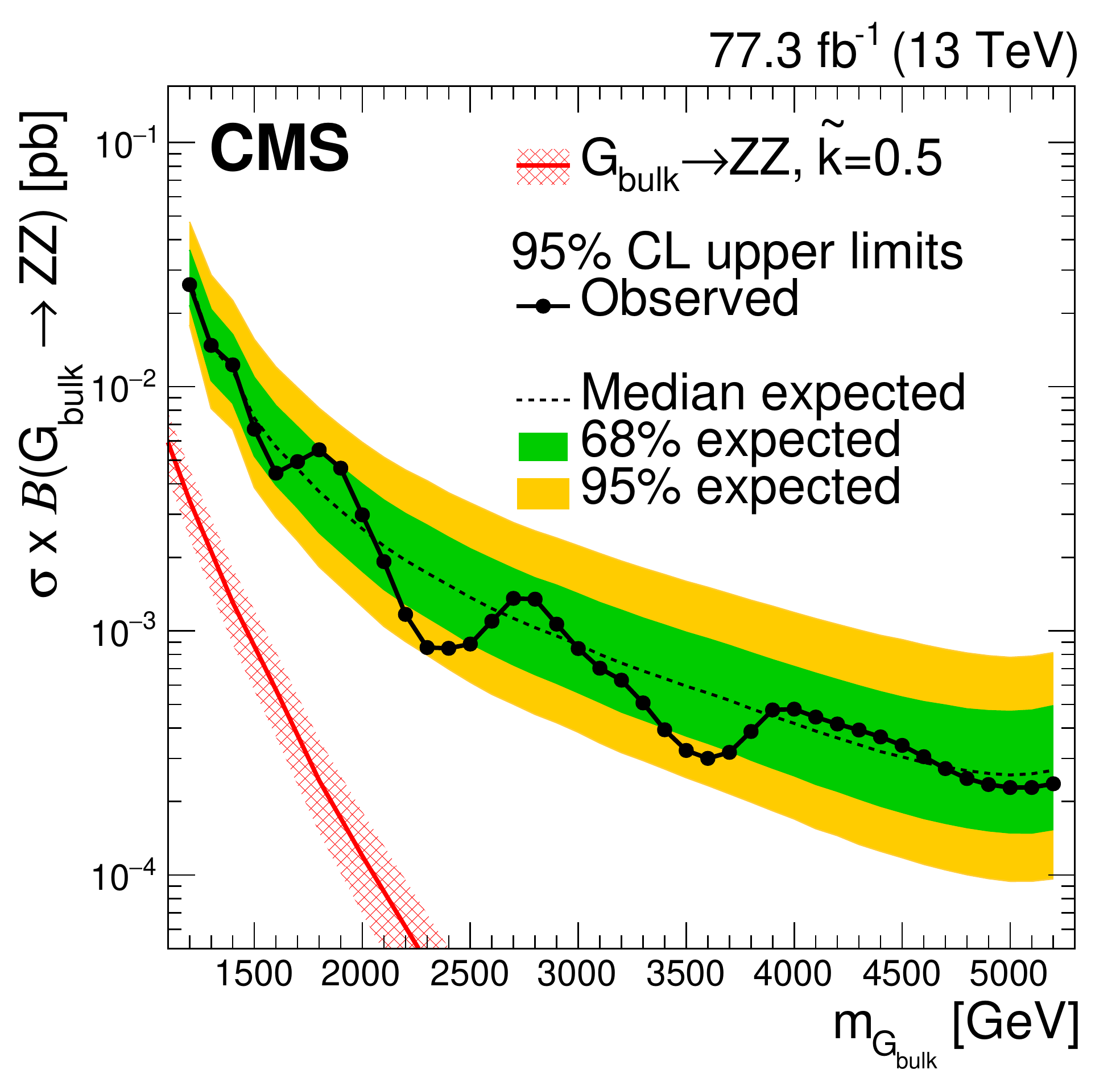}\\
\includegraphics[width=0.49\textwidth]{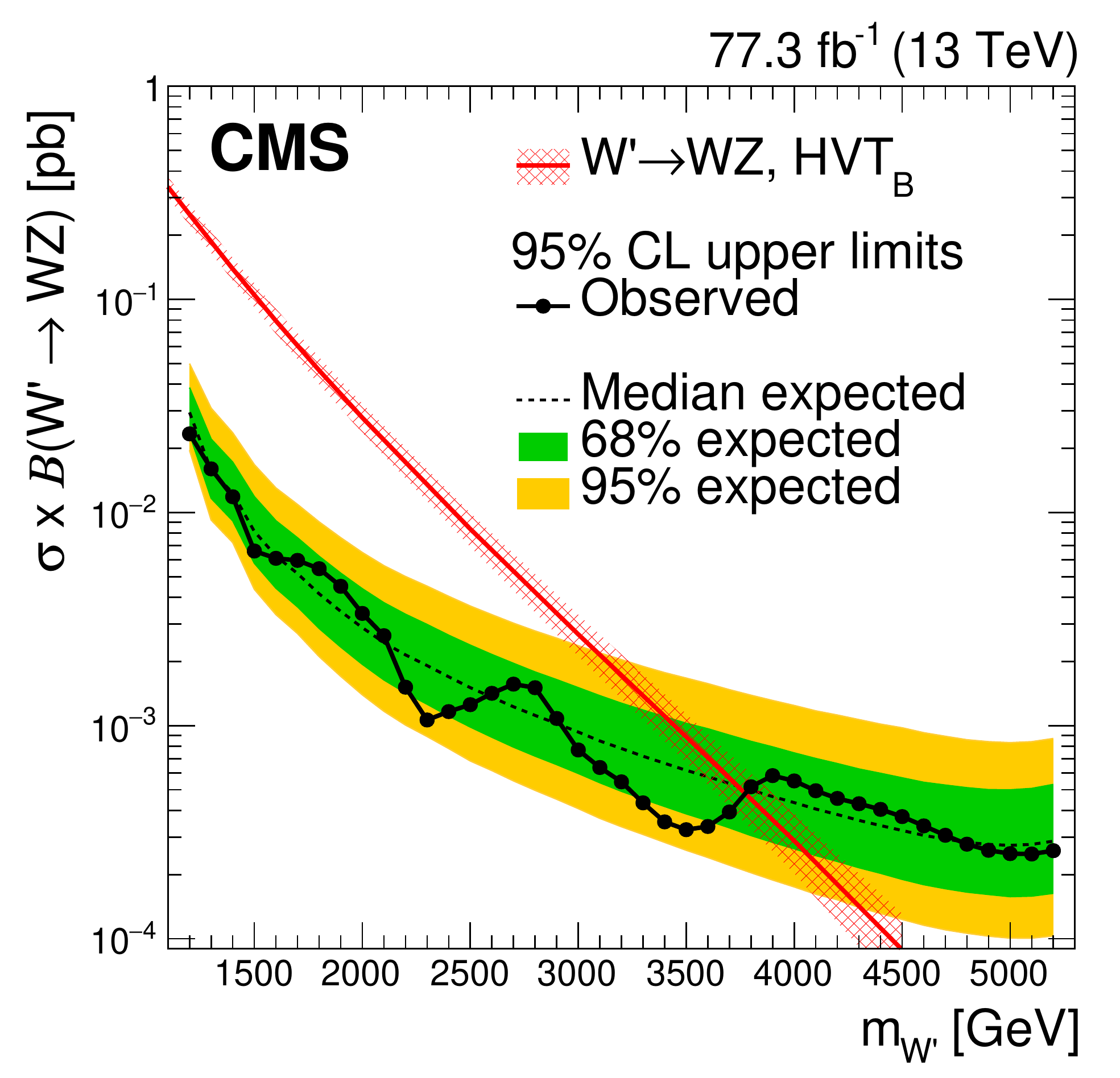}
\includegraphics[width=0.49\textwidth]{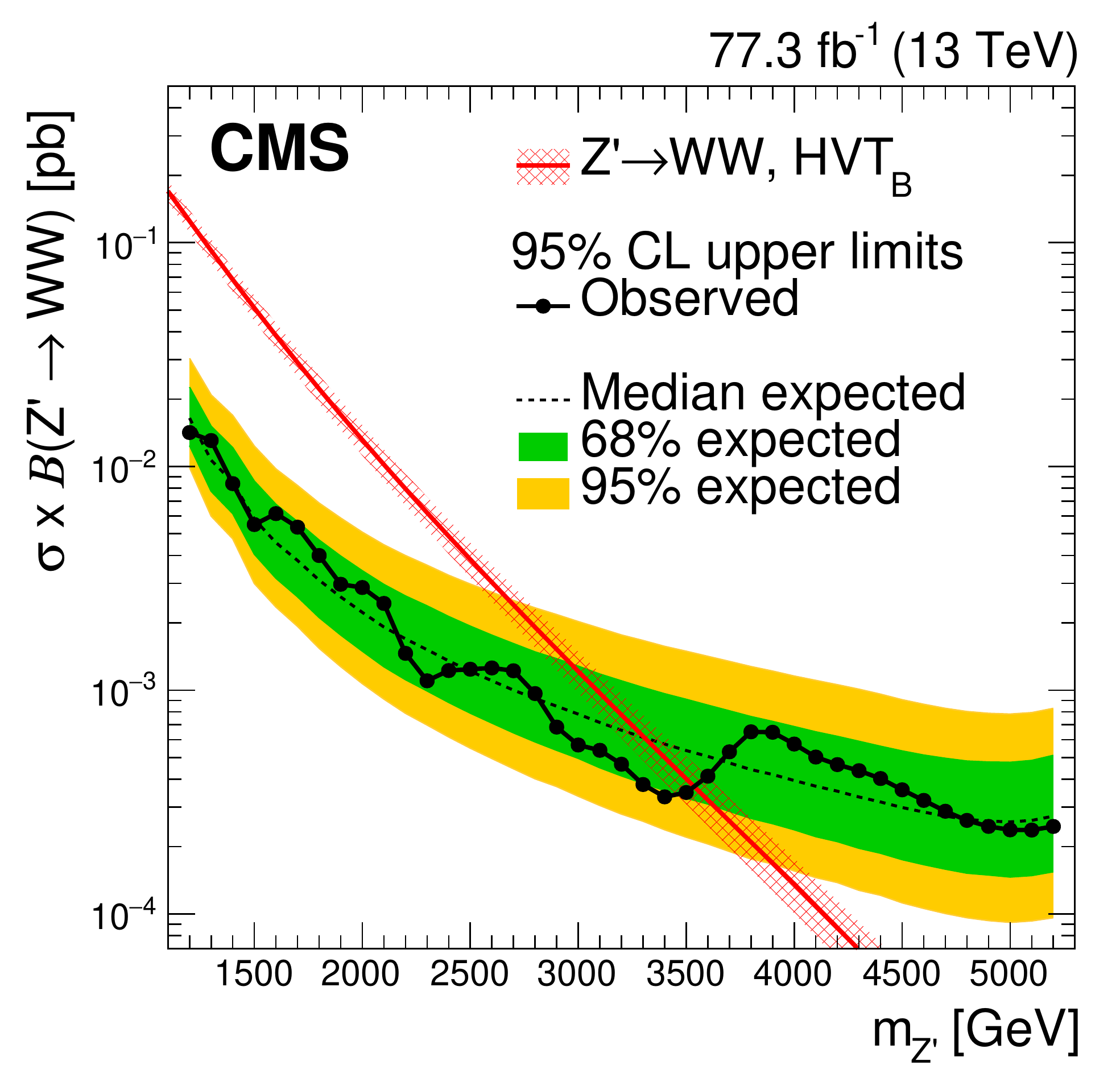}
\caption{Observed and expected 95\% \CL\ upper limits on the product of the production cross section ($\sigma$) and the branching fraction, obtained after combining categories of all purities with 77.3\fbinv of 13\TeV data, for $\BulkG\to\PW\PW$ (upper left), $\BulkG\to\PZ\PZ$  (upper right), $\PWpr\to\PW\PZ$ (lower left), and $\PZpr\to\PW\PW$ (lower right) signals. For each signal scenario the theoretical prediction (red line) and its uncertainty associated with the choice of PDF set (red hashed band) is shown. The theory cross sections (red line) are calculated at LO in QCD~\cite{Oliveira:2014kla,Pappadopulo:2014qza}.}
\label{fig:limits}
\end{figure*}

No significant excess over the background estimation is observed. Upper limits on the production cross section at 95\% confidence level (\CL) are set. Limits are set in the context of the bulk graviton model and the HVT model~B scenario, using the narrow-width approximation. Figure~\ref{fig:limits} shows the resulting limits as a function of the resonance mass compared to theoretical predictions. The theoretical cross sections shown in Figure~\ref{fig:limits} are calculated to LO in QCD as detailed in Ref~\cite{Oliveira:2014kla,Pappadopulo:2014qza}. For the HVT model~B, we exclude at 95\% \CL\ \PWpr and \PZpr spin-1 resonances with masses below \MassExclWPr and \MassExclZPr{}\TeV, respectively. In the narrow-width bulk graviton model, upper limits on the production cross sections for $\BulkG\to\PW{}\PW (\PZ{}\PZ{})$ are set in the range from \BulkGWWMassMinXsec{} (\BulkGZZMassMinXsec{})\unit{fb} for a resonance mass of \BulkGMassMin{}\TeV, down to \BulkGWWMassMaxXsec{}\unit{fb} for a resonance mass of \BulkGMassMax{}\TeV.

\begin{figure}[htbp]
\centering
\includegraphics[width=0.49\textwidth]{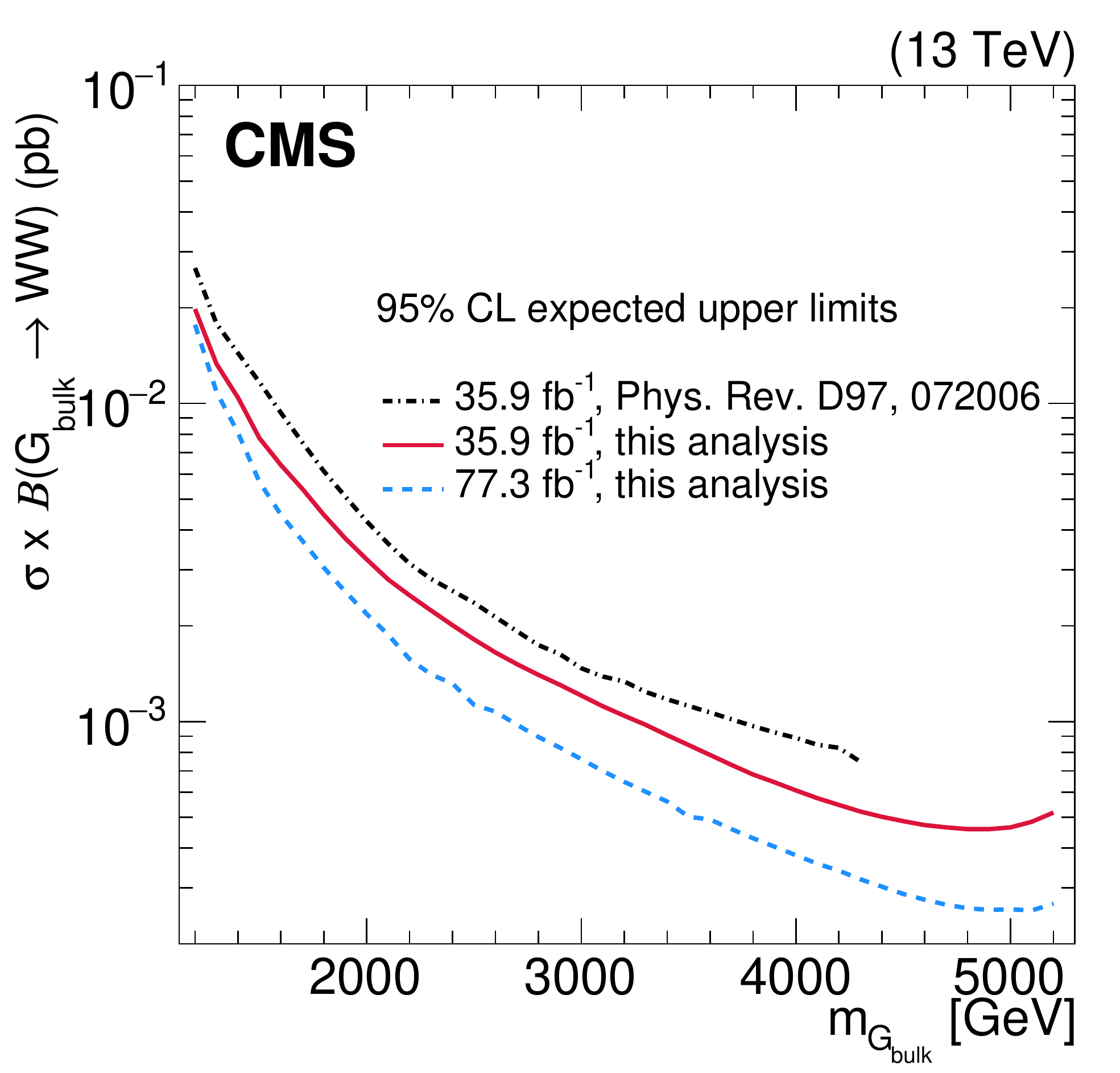}
\caption{Expected 95\% \CL\ upper limits on the product of the production cross section ($\sigma$) and the branching fraction for a $\BulkG\to\PW\PW$ signal using 35.9\fbinv of data collected in 2016 obtained using the multi-dimensional fit method presented here (red solid line), compared to the result obtained with previous methods (black dash-dotted line)~\cite{Sirunyan:2017acf}. The final limit obtained when combining data collected in 2016 and 2017 is also shown (blue dashed line).}
\label{fig:limitsCompare}
\end{figure}

The expected upper limits obtained using the multi-dimensional fit method introduced here are compared to those obtained in a previous search~\cite{Sirunyan:2017acf} using the same data set in order to estimate the sensitivity gain by using the new method.
Figure~\ref{fig:limitsCompare} shows the expected limits for one signal model based on an analyses of the data collected in 2016, using either the fit method presented here, or previous methods. We obtain a 20--30\% improvement in sensitivity when using the multi-dimensional fit method, and about a 35--40\% improvement when combining the data sets recorded in 2016 and 2017 relative to the individual results. The same conclusion holds for the other signal models investigated in this paper. The results obtained with the multi-dimensional fit are the best to date in the \PV{}\PV channel and reach a similar sensitivity at high masses (\MX) as the combination of diboson and leptonic decay channels for the 36\fbinv recorded in 2016~\cite{Sirunyan:2019vgt,Aaboud:2018bun}.

\section{Summary}
\label{sec:summary}

A search is presented for resonances with masses above \BulkGMassMin{}\TeV that decay to \PW{}\PW, \PZ{}\PZ, or \PW{}\PZ boson pairs. Each of the two bosons decays into one large-radius jet, yielding dijet final states. The search is conducted using a novel approach based on a three-dimensional maximum likelihood fit in the dijet invariant mass as well as the two jet masses, thus taking advantage of the fact that the expected signal is resonant in all three mass dimensions.
This method yields an improvement in sensitivity of up to 30\% relative to previous search methods.
The new method places additional constraints on systematic uncertainties affecting the signal by measuring the standard model background from \PW or \PZ production with associated jets, accounting for 5\% of the overall improvement in sensitivity.
Decays of $\PW$ and $\PZ$ bosons are identified using jet substructure observables that reduce the background from quantum chromodynamics multijet production.
No evidence is found for a signal, and upper limits on the resonance production cross section are set as a function of the resonance mass. The limits presented in this paper are the best to date in the dijet final state, and have a similar sensitivity as the combinations of diboson and leptonic decay channels using the 2016 data set. The results are interpreted within bulk graviton models, and as limits on the production of the \PWpr{} and \PZpr{} bosons within the heavy vector triplet framework.
For the heavy vector triplet model~B, we exclude at 95\% confidence level \PWpr and \PZpr spin-1 resonances with masses below \MassExclWPr and \MassExclZPr{}\TeV, respectively.
In the narrow-width bulk graviton model, upper limits on the production cross sections for $\BulkG\to\PW\PW~(\PZ{}\PZ{})$ are set in the range of \BulkGWWMassMinXsec{} (\BulkGZZMassMinXsec)\unit{fb} to \BulkGWWMassMaxXsec{}\unit{fb} for resonance masses between \BulkGMassMin{} and \BulkGMassMax{}\TeV.

\ifthenelse{\boolean{cms@external}}{\clearpage}{}

\begin{acknowledgments}
We congratulate our colleagues in the CERN accelerator departments for the excellent performance of the LHC and thank the technical and administrative staffs at CERN and at other CMS institutes for their contributions to the success of the CMS effort. In addition, we gratefully acknowledge the computing centres and personnel of the Worldwide LHC Computing Grid for delivering so effectively the computing infrastructure essential to our analyses. Finally, we acknowledge the enduring support for the construction and operation of the LHC and the CMS detector provided by the following funding agencies: BMBWF and FWF (Austria); FNRS and FWO (Belgium); CNPq, CAPES, FAPERJ, FAPERGS, and FAPESP (Brazil); MES (Bulgaria); CERN; CAS, MoST, and NSFC (China); COLCIENCIAS (Colombia); MSES and CSF (Croatia); RPF (Cyprus); SENESCYT (Ecuador); MoER, ERC IUT, PUT and ERDF (Estonia); Academy of Finland, MEC, and HIP (Finland); CEA and CNRS/IN2P3 (France); BMBF, DFG, and HGF (Germany); GSRT (Greece); NKFIA (Hungary); DAE and DST (India); IPM (Iran); SFI (Ireland); INFN (Italy); MSIP and NRF (Republic of Korea); MES (Latvia); LAS (Lithuania); MOE and UM (Malaysia); BUAP, CINVESTAV, CONACYT, LNS, SEP, and UASLP-FAI (Mexico); MOS (Montenegro); MBIE (New Zealand); PAEC (Pakistan); MSHE and NSC (Poland); FCT (Portugal); JINR (Dubna); MON, RosAtom, RAS, RFBR, and NRC KI (Russia); MESTD (Serbia); SEIDI, CPAN, PCTI, and FEDER (Spain); MOSTR (Sri Lanka); Swiss Funding Agencies (Switzerland); MST (Taipei); ThEPCenter, IPST, STAR, and NSTDA (Thailand); TUBITAK and TAEK (Turkey); NASU and SFFR (Ukraine); STFC (United Kingdom); DOE and NSF (USA).

\hyphenation{Rachada-pisek} Individuals have received support from the Marie-Curie programme and the European Research Council and Horizon 2020 Grant, contract Nos.\ 675440, 752730, and 765710 (European Union); the Leventis Foundation; the A.P.\ Sloan Foundation; the Alexander von Humboldt Foundation; the Belgian Federal Science Policy Office; the Fonds pour la Formation \`a la Recherche dans l'Industrie et dans l'Agriculture (FRIA-Belgium); the Agentschap voor Innovatie door Wetenschap en Technologie (IWT-Belgium); the F.R.S.-FNRS and FWO (Belgium) under the ``Excellence of Science -- EOS" -- be.h project n.\ 30820817; the Beijing Municipal Science \& Technology Commission, No. Z181100004218003; the Ministry of Education, Youth and Sports (MEYS) of the Czech Republic; the Lend\"ulet (``Momentum") Programme and the J\'anos Bolyai Research Scholarship of the Hungarian Academy of Sciences, the New National Excellence Program \'UNKP, the NKFIA research grants 123842, 123959, 124845, 124850, 125105, 128713, 128786, and 129058 (Hungary); the Council of Science and Industrial Research, India; the HOMING PLUS programme of the Foundation for Polish Science, cofinanced from European Union, Regional Development Fund, the Mobility Plus programme of the Ministry of Science and Higher Education, the National Science Center (Poland), contracts Harmonia 2014/14/M/ST2/00428, Opus 2014/13/B/ST2/02543, 2014/15/B/ST2/03998, and 2015/19/B/ST2/02861, Sonata-bis 2012/07/E/ST2/01406; the National Priorities Research Program by Qatar National Research Fund; the Ministry of Science and Education, grant no. 3.2989.2017 (Russia); the Programa Estatal de Fomento de la Investigaci{\'o}n Cient{\'i}fica y T{\'e}cnica de Excelencia Mar\'{\i}a de Maeztu, grant MDM-2015-0509 and the Programa Severo Ochoa del Principado de Asturias; the Thalis and Aristeia programmes cofinanced by EU-ESF and the Greek NSRF; the Rachadapisek Sompot Fund for Postdoctoral Fellowship, Chulalongkorn University and the Chulalongkorn Academic into Its 2nd Century Project Advancement Project (Thailand); the Welch Foundation, contract C-1845; and the Weston Havens Foundation (USA).
\end{acknowledgments}

\bibliography{auto_generated}
\cleardoublepage \appendix\section{The CMS Collaboration \label{app:collab}}\begin{sloppypar}\hyphenpenalty=5000\widowpenalty=500\clubpenalty=5000\vskip\cmsinstskip
\textbf{Yerevan Physics Institute, Yerevan, Armenia}\\*[0pt]
A.M.~Sirunyan$^{\textrm{\dag}}$, A.~Tumasyan
\vskip\cmsinstskip
\textbf{Institut f\"{u}r Hochenergiephysik, Wien, Austria}\\*[0pt]
W.~Adam, F.~Ambrogi, T.~Bergauer, J.~Brandstetter, M.~Dragicevic, J.~Er\"{o}, A.~Escalante~Del~Valle, M.~Flechl, R.~Fr\"{u}hwirth\cmsAuthorMark{1}, M.~Jeitler\cmsAuthorMark{1}, N.~Krammer, I.~Kr\"{a}tschmer, D.~Liko, T.~Madlener, I.~Mikulec, N.~Rad, J.~Schieck\cmsAuthorMark{1}, R.~Sch\"{o}fbeck, M.~Spanring, D.~Spitzbart, W.~Waltenberger, C.-E.~Wulz\cmsAuthorMark{1}, M.~Zarucki
% ...
\vskip\cmsinstskip
\dag: Deceased\\
1:  Also at Vienna University of Technology, Vienna, Austria\\
% ...
\end{sloppypar}
\end{document}